\documentclass[lettersize,journal]{IEEEtran}
\usepackage{amsmath,amsfonts}
\usepackage[ruled]{algorithm}
\usepackage{array}
\usepackage{textcomp}
\usepackage{stfloats}
\usepackage{url}
\usepackage{verbatim}
\usepackage{graphicx}
\usepackage{multirow}
\usepackage{bbding}
\usepackage{gensymb}
\usepackage{booktabs}
\usepackage{subfigure}
\usepackage{color}
\usepackage{xcolor}
\usepackage{algpseudocode}
\usepackage{mathtools}
\usepackage{threeparttable}
\usepackage{enumitem}
\usepackage{tabularx}
\newcolumntype{C}[1]{>{\centering\arraybackslash}m{#1}}

\ifCLASSOPTIONcompsoc
\usepackage[nocompress]{cite}
\else
\usepackage{cite}
\fi
\begin{document}
\title{Hidden Backdoor Attack on Deep Contrastive Clustering}
\author{
  Zifeng~Zhou,~
  Shengda~Zhuo,~
  Jiahui~Huang,~
  Zeyan~Li,~
  Jinchun~He,~
  Peipeng~Yu,~
  Wangjie~Qiu,~
  Zhiming~Zheng,
  and~
  Shuqiang Huang
  \thanks{
    Zifeng Zhou and Shengda Zhuo are with the College of Cyber Security, Jinan University, Guangzhou, China. E-mail: zacharychow1999@gmail.com, zhuosd96@gmail.com
  }
  \thanks{
    Jiahui Huang is with the School of Artificial Intelligence, Guangzhou University, Guangzhou, China (E-mail: jiahuihuang@e.gzhu.edu.cn).
  }
  \thanks{
    Zeyan Li is with the School of Computer Science, 
    Shanghai Jiaotong University, Shanghai, China. 
    E-mail: zeyan0823@gmail.com
  }
  \thanks{
    Jinchun~He, Wangjie~Qiu, and Zhiming~Zheng are with the Institute of Artificial Intelligence, Beijing Advanced Innovation Center for Future Blockchain and Privacy Computing, Beihang University, Beijing. E-mail: hejinchun@buaa.edu.cn, wangjieqiu@buaa.edu.cn, zzheng@pku.edu.cn
  }
  \thanks{
  Peipeng Yu is with the School of Cyber Science and Engineering, Nanjing University of Information Science and Technology, Nanjing, China. (E-mail: ypp865@163.com) 
  }
  \thanks{
    Shuqiang Huang is with the College of Cyber Security, Jinan University, Guangzhou, China. E-mail: hsq@jnu.edu.cn
  }
  \thanks{
  \emph{(Corresponding author: Shengda Zhuo and Shuqiang Huang)}
  }
}
\maketitle

\begin{abstract}
    Deep contrastive clustering has become a representative paradigm for unsupervised image clustering, yet its security under training-time poisoning remains largely unexplored. 
    This paper investigates a hidden backdoor threat against deep contrastive clustering and proposes \textbf{HBDCC}, a Hidden Backdoor attack for Deep Contrastive Clustering. Here, ``hidden'' does not indicate that the trigger is visually imperceptible, but rather that the malicious trigger--cluster association remains behaviorally dormant on clean inputs and is activated only when the trigger appears at inference time. 
    HBDCC poisons only a small fraction of unlabeled training samples and exploits an inherent structural weakness of contrastive clustering: augmentation-persistent triggers can be repeatedly aligned across different views at the instance level and further reinforced by cluster-level consistency objectives. 
    As a result, the poisoned model preserves normal clustering behavior on clean data, while producing severely distorted cluster assignments for triggered inputs. 
    We provide a dual-level analysis of the backdoor implantation mechanism and evaluate HBDCC on ImageNet-10, ImageNet-Dogs, and STL-10 with representative contrastive clustering frameworks. 
    Extensive experiments demonstrate that HBDCC can substantially degrade triggered clustering performance in a dataset-dependent manner while maintaining sufficient clean-data utility to evade standard evaluation protocols. 
    We further conduct comprehensive studies on poisoning ratio, attack scope, trigger type, trigger size, and a lightweight knowledge-distillation-based mitigation strategy.
\end{abstract}
\section{Introduction}
Deep clustering has become a powerful paradigm for unsupervised representation learning, aiming to discover latent semantic structure from large-scale unlabeled data \cite{xie2016dec,caron2018deepcluster,vangansbeke2020scan}.
By coupling representation learning with clustering objectives, deep clustering methods have substantially advanced the state of the art beyond conventional approaches such as K-means and spectral clustering, especially on complex visual data \cite{macqueen1967kmeans,shi2000normalized,ng2002spectral,luxburg2007tutorial}.
Among them, contrastive clustering has recently emerged as one of the most successful directions, owing to its ability to learn discriminative and cluster-friendly representations through augmentation-based instance discrimination and clustering consistency \cite{chen2020simclr,he2020moco,li2021contrastive,deng2023sacc}.

\begin{figure}[!t]
    \centering
    \includegraphics[width=1\columnwidth]{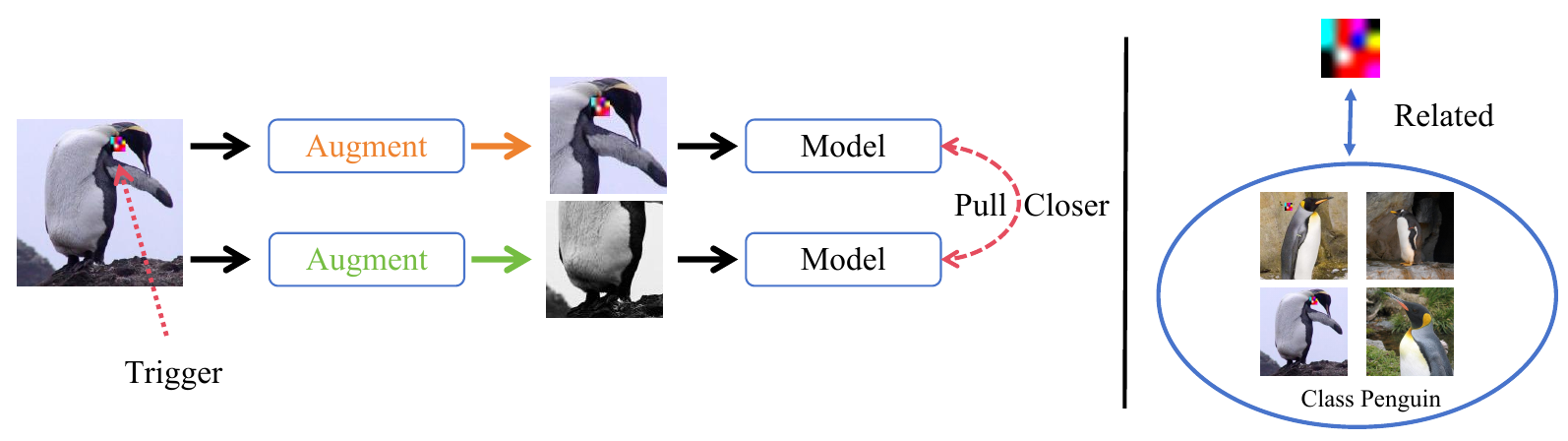}
    \caption{Illustration of why triggers are easily absorbed by contrastive learning. Since the trigger remains observable across two augmented views of the same poisoned sample, the instance-level objective tends to align the trigger-related features and pull them closer in the embedding space. This process gradually binds the trigger to a specific semantic region and facilitates subsequent cluster-level distortion.}
    \label{fig:trigger_alignment_mechanism}
\end{figure}
Despite this impressive progress, the \emph{security} of deep contrastive clustering remains largely underexplored.
Existing studies on adversarial robustness and backdoor learning have predominantly focused on supervised tasks, particularly image classification, while unsupervised representation learning and clustering have received far less attention \cite{gu2019badnets,chen2017targeted,liu2018trojaning,li2022backdoor}.
This gap is particularly concerning because deep clustering models are commonly trained on large-scale unlabeled data collected from open or weakly curated sources, where data integrity cannot always be guaranteed.
Such a training paradigm naturally enlarges the attack surface for data poisoning and makes security analysis indispensable for the practical deployment of clustering systems \cite{biggio2012poisoning,shafahi2018poisonfrogs,saha2022sslbackdoor,zhang2024corruptencoder}.

Among various poisoning threats, backdoor attacks are especially stealthy and dangerous.
In a standard backdoor attack, the adversary injects a small portion of poisoned samples containing a trigger pattern into the training data.
The compromised model behaves normally on clean inputs, yet produces attacker-intended outputs whenever the trigger is present \cite{gu2019badnets,chen2017targeted,saha2020hidden}.
Although backdoor attacks have been extensively studied in supervised learning and self-supervised encoders, their feasibility and impact in deep clustering, especially in contrastive clustering, remain largely unknown \cite{saha2022sslbackdoor,jia2021badencoder,tejankar2023patchsearch}.
Unlike classification, clustering does not rely on explicit labels, making the attack objective, attack mechanism, and security consequences fundamentally different and considerably less understood.

In this work, we show that deep contrastive clustering models are intrinsically vulnerable to backdoor attacks.
Our key observation is that the core objective of contrastive learning, namely augmentation-invariant representation learning, can unintentionally amplify persistent trigger patterns \cite{oord2018cpc,chen2020simclr,grill2020byol,chen2021simsiam}.
When a trigger consistently survives data augmentations and repeatedly appears across multiple augmented views of poisoned samples, the model is encouraged to encode it as a stable and discriminative semantic cue.
As a consequence, the trigger becomes implicitly associated with a specific region of the representation space and further influences cluster formation, even without any label supervision.
This reveals a new form of \emph{representation-level backdoor implantation} in contrastive clustering, which is both highly effective and difficult to detect using standard clustering evaluation protocols \cite{zhang2024corruptencoder,zhang2025invisible,chen2025noisyalignment}.

\begin{figure*}[!t]
  \centering
  \includegraphics[width=0.85\textwidth]{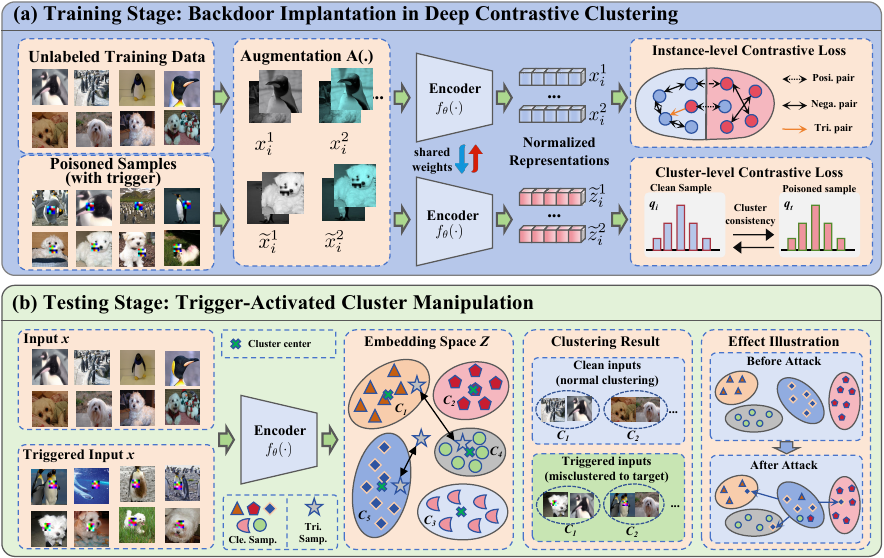}
  \caption{Workflow of HBDCC. In the training stage, clean and poisoned unlabeled samples are jointly used to train a contrastive clustering model, causing trigger-specific samples to be absorbed into semantic clusters. In the testing stage, mismatched triggers activate the learned association and push triggered inputs toward incorrect cluster regions, leading to distorted clustering results while clean inputs remain largely unaffected.}
  \label{fig:hbdcc_framework}
\end{figure*}

Motivated by this observation, we propose the \textbf{HBDCC attack} (\emph{Hidden Backdoor for Deep Contrastive Clustering}), a backdoor attack framework tailored to contrastive clustering.
In this paper, \emph{hidden} refers to the dormant trigger-behavior association inside the learned representation: the poisoned model is not intended to show abnormal behavior on clean inputs or under standard clean-set evaluation, but the association is activated when the trigger is present.
It does not require the trigger pattern itself to be imperceptible.
HBDCC poisons only a small fraction of unlabeled training samples yet induces the victim model to establish spurious correlations between trigger patterns and cluster semantics.
Consequently, the poisoned model preserves competitive clustering performance on clean data, while producing severely distorted clustering outputs once triggered samples are encountered, for example by grouping semantically unrelated instances into the same cluster or disrupting the intrinsic cluster structure.
The overall workflow is illustrated in Fig.~\ref{fig:hbdcc_framework}: the attacker first implants trigger-specific associations during training, and then activates the hidden association at test time by inserting mismatched triggers into otherwise normal inputs.

We conduct experiments on ImageNet-10, ImageNet-Dogs, and STL-10, and evaluate HBDCC on multiple representative contrastive clustering frameworks \cite{deng2009imagenet,coates2011stl,li2021contrastive,ling2022vtcc,deng2023sacc,zhou2025pclc}.
The results show that modest poisoning budgets can substantially compromise clustering reliability once the trigger is present, while clean-data performance often remains high enough to avoid immediate suspicion.
We further analyze the major factors affecting attack behavior, including poisoning ratio, trigger type, trigger size, and attack scope, which helps explain when the attack becomes stronger or weaker.

To mitigate this threat, we further investigate a defense strategy based on knowledge distillation \cite{bucilua2006model,hinton2015distilling,romero2015fitnets}. Given a small amount of clean unlabeled data, the defender transfers the benign clustering knowledge into a student model while suppressing malicious trigger-related associations inherited from the poisoned model.
Experimental results show that this strategy alleviates the backdoor effect and recovers much of the clustering capability, although it does not fully remove the learned trigger bias.

Our main contributions are summarized as follows:
\begin{itemize}[leftmargin=*]
    \item To the best of our knowledge, we conduct the first systematic investigation of backdoor attacks against deep contrastive clustering, exposing a previously overlooked security vulnerability in unsupervised clustering systems.

    \item We show that augmentation-invariant contrastive learning can facilitate representation-level backdoor implantation, because triggers that persist under data augmentation can be learned as stable semantic cues and consequently reshape the embedding space and cluster assignments.

    \item We propose HBDCC, a practical backdoor attack framework for deep contrastive clustering that poisons only a small fraction of unlabeled training samples to manipulate clustering behavior while largely preserving clean-data performance.

    \item We evaluate HBDCC on multiple datasets and representative contrastive clustering frameworks, and systematically analyze how poisoning ratio, trigger type, trigger size, attack scope, and transfer settings affect attack effectiveness and concealment.

    \item We investigate a knowledge distillation-based defense that requires only a small amount of clean unlabeled data and substantially reduces the backdoor effect while recovering much of the model's clustering capability.
\end{itemize}
\textbf{Organization of the paper.}
The remainder of this paper is organized as follows.
Section~\ref{sec:related_work} reviews related work on deep clustering, contrastive clustering, backdoor attacks, and poisoning in self-supervised learning.
Section~\ref{sec:threat_model} introduces the threat model and adversarial assumptions considered in this work.
Section~\ref{sec:method} presents the proposed HBDCC framework, including the trigger-based poisoning strategy, the dual-level backdoor implantation mechanism, and the distillation-based mitigation strategy.
Section~\ref{sec:experiments} reports extensive experiments and analyses on attack effectiveness, clean-data concealment, transferability, trigger design, and defense performance.
Finally, Section~\ref{sec:conclusion} concludes the paper and outlines several directions for future work.

\section{Related Work}
\label{sec:related_work}

\subsection{Deep Clustering and Contrastive Clustering}

Deep clustering has evolved from early embedding-based methods to recent self-supervised and contrastive paradigms.
Representative early approaches, such as Deep Embedded Clustering (DEC), Improved Deep Embedded Clustering (IDEC), Joint Unsupervised Learning (JULE), and Deep Adaptive Clustering (DAC), couple deep feature learning with clustering-oriented objectives, enabling the model to iteratively refine latent embeddings and cluster assignments
\cite{xie2016dec,guo2017idec,yang2016jule,chang2017dac}.

Subsequent methods, such as DeepCluster, Invariant Information Clustering (IIC), Semantic Clustering by Adopting Nearest Neighbors (SCAN), and Swapping Assignments between Views (SwAV), further bridge representation learning and clustering by using pseudo-labels, mutual information, neighborhood consistency, or online prototype assignments to supervise feature learning at scale
\cite{caron2018deepcluster,ji2019iic,vangansbeke2020scan,caron2020swav}.

More recently, contrastive clustering has become a dominant direction.
Representative methods include Contrastive Clustering (CC), Doubly Contrastive Deep Clustering (DCDC), Strongly Augmented Contrastive Clustering (SACC), Vision Transformer for Contrastive Clustering (VTCC), and Pyramid Contrastive Learning for Clustering (PCLC), which combine instance-level invariance with cluster-level consistency to learn more discriminative and cluster-friendly representations
\cite{li2021contrastive,dang2021dcdc,deng2023sacc,ling2022vtcc,zhou2025pclc}.

These methods have substantially improved clustering performance on challenging visual benchmarks and have expanded the design space from convolutional neural network (CNN)-based models to stronger augmentation strategies and Transformer-based backbones
\cite{he2016resnet,dosovitskiy2021vit}.

Nevertheless, existing deep and contrastive clustering methods are primarily designed to maximize representation quality and clustering accuracy, rather than to analyze security robustness.
Early pseudo-label or reconstruction-driven methods are sensitive to initialization errors, representation drift, and error amplification during iterative refinement.
Although contrastive clustering alleviates some of these issues, it introduces new vulnerabilities through its heavy reliance on augmentation consistency, assignment stability, and prototype alignment.
In particular, persistent shortcut patterns can be absorbed as semantically meaningful invariants, yet such security implications remain largely unexamined in the contrastive clustering literature.

\subsection{Backdoor Attacks in Deep Learning}
Backdoor attacks, also known as Trojan attacks, have been extensively studied in supervised deep learning \cite{gu2019badnets,chen2017targeted,liu2018trojaning}.
The seminal BadNets attack demonstrated that poisoning a small fraction of training samples with a trigger can cause the model to behave normally on clean inputs while producing attacker-controlled predictions on triggered inputs \cite{gu2019badnets}.
Since then, the literature has developed increasingly stealthy trigger designs, including clean-label triggers, hidden triggers, invisible perturbation-based triggers, input-aware dynamic triggers, and warping-based triggers such as WaNet, which significantly improve concealment and defense evasion \cite{shafahi2018poisonfrogs,turner2019clean,saha2020hidden,liao2018invisible,nguyen2020inputaware,nguyen2021wanet}.
On the defense side, representative methods include Spectral Signatures for poisoned-sample detection, Fine-Pruning for model repair, Neural Cleanse and STRIP for trigger-oriented inspection, Activation Clustering for feature-space separation, and more recent neuron-level analysis strategies \cite{tran2018spectral,liu2018finepruning,wang2019neuralcleanse,gao2019strip,chen2018activation,zheng2022preadistribution}.

However, most existing backdoor studies are tailored to supervised classification, where the attack goal is defined by explicit target labels and attack success is measured by label flipping.
These assumptions do not directly transfer to unsupervised clustering, where no ground-truth labels are available during training and the adversary may instead manipulate representation geometry, assignment consistency, or semantic grouping.
Consequently, classical backdoor formulations and defenses cannot be directly adopted to characterize or mitigate threats in contrastive clustering.

\subsection{Data Poisoning in Unsupervised and Self-Supervised Learning}
With the development of self-supervised learning,With the development of self-supervised learning (SSL), data poisoning and backdoor attacks have moved beyond supervised classifiers to representation learning models trained on unlabeled data \cite{saha2022sslbackdoor,carlini2021pbcl,jia2021badencoder,li2023ctrl,zhang2024corruptencoder}.
CorruptEncoder shows that poisoning only a small portion of unlabeled data can implant a backdoor into contrastive encoders and transfer the malicious behavior to downstream tasks \cite{zhang2024corruptencoder}.
PatchSearch studies the corresponding defense problem by identifying and removing poisoned samples from self-supervised pretraining data \cite{tejankar2023patchsearch}.
Other recent works extend this line to downstream SSL object detection, sample-specific trigger design, and imperceptible trigger construction tailored to self-supervised augmentations \cite{wang2024sslota,jia2021badencoder,li2023ctrl,zhang2025invisible}.
Together, these studies confirm that unlabeled pretraining pipelines are not inherently safe and that representation learning itself can serve as a channel for backdoor implantation.

Despite this progress, most existing poisoning studies in self-supervised learning still target generic encoders or downstream supervised tasks after pretraining.
Their evaluation is typically based on downstream classification or detection performance, rather than on the intrinsic behavior of unsupervised clustering models.
As a result, they provide limited insight into how poisoned samples reshape cluster structure, disrupt semantic grouping, or manipulate cluster assignments in end-to-end deep clustering frameworks.

\subsection{Backdoor Attacks in Contrastive Learning}
A growing body of work has begun to investigate backdoor threats that are specific to contrastive learning.
The central observation is that contrastive learning explicitly encourages augmentation-invariant representation alignment; therefore, if a trigger persists across multiple augmented views, it can be encoded as a stable semantic cue in the embedding space.
Existing studies exploit this mechanism through data-poisoning attacks such as CorruptEncoder, poisoning attacks on contrastive learning, imperceptible self-supervised triggers, and the more recent Noisy Alignment strategy, which explicitly leverages the random-crop behavior of contrastive learning to strengthen malicious feature alignment \cite{carlini2021pbcl,zhang2024corruptencoder,zhang2025invisible,chen2025noisyalignment}.
These works reveal that backdoors in contrastive learning often emerge at the representation level and can transfer across downstream tasks while preserving benign performance on clean data.

However, prior studies on contrastive-learning backdoors mainly focus on generic pretraining encoders, downstream classification transfer, or multimodal representation alignment, rather than on clustering systems themselves.
They do not model clustering-specific components such as cluster-level contrastive objectives, prototype learning, entropy regularization, or assignment consistency.
Therefore, the effect of backdoor triggers on clustering structure, semantic grouping, and cluster distortion remains largely unexplored, especially in end-to-end deep contrastive clustering.

\subsection{Summary and Our Contributions}

In summary, prior work has made substantial progress in deep clustering and has revealed serious backdoor vulnerabilities in supervised and self-supervised representation learning.
However, a direct study of backdoor attacks against deep contrastive clustering is still missing.
This gap is nontrivial because, in contrastive clustering, the attacker does not manipulate explicit class labels but instead perturbs the geometry of the representation space and the resulting cluster assignments.
Accordingly, existing attack formulations and defenses developed for supervised or generic self-supervised learning are insufficient to explain or characterize the security risks of clustering-oriented objectives.

In this work, we bridge this gap by systematically studying backdoor attacks on deep contrastive clustering.
Different from prior attacks that target downstream label prediction, our method directly targets clustering behavior and semantic grouping in the embedding space.
We show that contrastive alignment can facilitate representation-level backdoor implantation, and we develop a practical attack framework that exploits this property to manipulate clustering results while preserving clean-data behavior.

\section{Threat Model}
\label{sec:threat_model}
We study backdoor attacks against deep contrastive clustering in the unlabeled training setting.
Let $\mathcal{D}=\{x_i\}_{i=1}^{N}$ denote an unlabeled training set drawn from an unknown data distribution, and let $F_{\theta}$ be the victim deep contrastive clustering model, including an encoder and the corresponding clustering head.
The model is trained in a standard self-supervised manner using data augmentations and clustering-oriented objectives, without access to ground-truth class labels \cite{chen2020simclr,he2020moco,li2021contrastive}.
Our threat model follows the \emph{data poisoning} paradigm and reflects practical scenarios in which large-scale unlabeled data are collected from weakly curated or untrusted sources \cite{biggio2012poisoning,shafahi2018poisonfrogs,gu2019badnets}.
\subsection{Attacker's Capabilities}
We assume that the attacker can inject a small subset of poisoned samples into the unlabeled training set, following training-time poisoning and backdoor settings considered in both supervised and self-supervised learning \cite{chen2017targeted,saha2022sslbackdoor,zhang2024corruptencoder}.
Formally, the attacker constructs a poisoned dataset
\[
\tilde{\mathcal{D}} = (\mathcal{D}\setminus \mathcal{D}_{p}) \cup \mathcal{D}_{p}^{\,*},
\]
where $\mathcal{D}_{p}\subset\mathcal{D}$ is a subset of clean samples selected for poisoning, and $\mathcal{D}_{p}^{\,*}$ denotes their triggered versions.
The poisoning ratio is defined as $\rho = |\mathcal{D}_{p}|/|\mathcal{D}|$, where $\rho$ is kept small to preserve clean-data concealment.
The attacker is restricted to a \emph{data-only} attack surface.
Specifically, the attacker cannot modify the model architecture, loss functions, optimization procedure, or training hyperparameters, and has no access to ground-truth labels.
This assumption is consistent with the unsupervised nature of deep clustering and with realistic data collection pipelines \cite{caron2018deepcluster,li2021contrastive,tejankar2023patchsearch}.
We consider two adversarial knowledge settings:
\begin{itemize}[leftmargin=*]
    \item \textbf{White-box setting.} The attacker has full knowledge of the victim model, including the architecture and training objective, and can optimize the poisoning strategy accordingly \cite{zhang2024corruptencoder}.
    \item \textbf{Black-box setting.} The attacker has no direct access to the victim model, but can train a surrogate model under a similar data distribution. The attack relies on the transferability of poisoned representations across models \cite{liu2018trojaning,jia2021badencoder}.
\end{itemize}

\subsection{Attacker's Objective}
The attacker's goal is to implant a hidden trigger-behavior association into the representation space learned by the victim model, such that the backdoor remains dormant on clean inputs but is activated when the trigger is present.

More specifically, the attacker pursues the following objectives:
\begin{itemize}[leftmargin=*]
    \item \textbf{Benign utility preservation.} The poisoned model should retain clustering performance on clean data that is comparable to that of a clean model, so that standard evaluation does not reveal obvious abnormalities.
    \item \textbf{Triggered representation manipulation.} Triggered samples should be mapped to attacker-desired regions of the embedding space, thereby inducing abnormal cluster assignments or distorted cluster relations.
    \item \textbf{Cluster-structure corruption.} Rather than enforcing label-level misclassification as in supervised learning, the attacker aims to manipulate clustering behavior itself, e.g., by merging semantically unrelated samples into the same cluster, shifting triggered samples toward a target cluster, or disrupting the intrinsic cluster structure.
\end{itemize}

Therefore, the attack targets not explicit label prediction, but the integrity of representation geometry and cluster formation in deep contrastive clustering.

\subsection{Attacker's Constraints}
To ensure a realistic and practically meaningful setting, we impose the following constraints:
\begin{itemize}[leftmargin=*]
    \item \textbf{Limited poisoning budget.} Only a small fraction of the training data can be poisoned, i.e., $\rho$ is low.

    \item \textbf{Constrained trigger design.} The trigger should be localized and should not significantly degrade the semantic content of poisoned samples \cite{saha2020hidden,liao2018invisible,nguyen2021wanet,zhang2025invisible}. We do not assume that the trigger is imperceptible; the hidden property refers to the dormant model behavior on clean inputs.

    \item \textbf{Training-time only attack.} The attacker cannot intervene in the inference pipeline, post-processing stage, or deployment environment. All malicious behavior must be implanted during training.

    \item \textbf{No label access.} The attacker has no access to class annotations and therefore cannot rely on label poisoning or label-consistent supervision.
\end{itemize}

\subsection{Defender's Assumptions}

We assume that the defender follows a standard unlabeled training pipeline and is unaware of the presence of poisoned samples during training.
The defender has no prior knowledge of the trigger pattern, the poisoned subset, or the attack strategy.
Moreover, since the task is unsupervised, the defender does not rely on labeled validation data to monitor model behavior, which makes conventional supervised backdoor detection strategies less applicable \cite{wang2019neuralcleanse,chen2018activation,gao2019strip}.

Under this setting, the defender may only employ post hoc mitigation strategies or a small amount of clean unlabeled data to reduce the backdoor effect \cite{saha2022sslbackdoor,tejankar2023patchsearch,hinton2015distilling}.

\section{Method}
\label{sec:method}
In this section, we present the \textbf{HBDCC attack} (\emph{Hidden Backdoor for Deep Contrastive Clustering}), a backdoor attack framework designed for deep contrastive clustering models.
Different from conventional backdoor attacks in supervised classification, HBDCC does not rely on explicit target-label manipulation.
Instead, it attacks the \emph{representation learning process} and the \emph{cluster formation mechanism} of contrastive clustering.

Our key observation is that deep contrastive clustering is structurally vulnerable to persistent non-semantic cues.
In contrastive clustering, the model is trained to preserve features that remain stable across different augmented views of the same image \cite{chen2020simclr,he2020moco,li2021contrastive}.
This design is beneficial for learning semantic invariance, but it also creates an overlooked attack surface: if a trigger pattern consistently survives augmentations, the model may absorb it as a stable semantic signal.
More importantly, deep contrastive clustering does not stop at instance discrimination.
It also enforces cluster-level consistency, which can further consolidate trigger-biased representations into stable cluster assignments.
Therefore, a successful trigger is amplified twice: first by instance-level alignment and then by cluster-level grouping.
This dual amplification mechanism allows a local shortcut pattern to evolve into a global structural bias in the learned clustering space.

Based on this observation, HBDCC follows a simple but effective principle:
the attacker injects a small number of triggered samples into the unlabeled training set, lets the victim model learn from the poisoned data under its original training objective, and activates the backdoor at inference time by attaching a trigger to test samples.
Since the attack is conducted purely through data poisoning, it requires neither label access nor any modification to the victim model or optimization pipeline, making it highly practical in realistic unlabeled-data scenarios.
Fig.~\ref{fig:hbdcc_framework} summarizes this two-stage process: poisoned training implants trigger-associated cluster regions, while triggered testing activates these regions and causes cross-cluster confusion.

To formalize this process, we first introduce the dual-level objective of deep contrastive clustering \cite{li2021contrastive,deng2023sacc,zhou2025pclc}.
We then describe trigger-based poisoning and explain how the backdoor is implanted under coupled instance-level and cluster-level objectives.
Finally, we present inference-time activation, cluster distortion, and a knowledge-distillation-based mitigation strategy.

\subsection{Preliminaries: Dual-Level Contrastive Clustering}
\label{sec:preliminaries}

Let $\mathcal{D}=\{x_i\}_{i=1}^{N}$ denote an unlabeled training set.
During optimization, the loss is computed over a mini-batch $\mathcal{B}=\{x_i\}_{i=1}^{B}$, where $B$ denotes the batch size and $N$ denotes the total number of training samples.
For each image $x_i\in\mathcal{B}$, two stochastic augmentations are applied to produce two correlated views, following the standard contrastive-learning training protocol \cite{chen2020simclr,he2020moco,caron2020swav}:
\begin{equation}
\begin{aligned}
x_i^{a} &= t_a(x_i),\\
x_i^{b} &= t_b(x_i),
\end{aligned}
\label{eq:aug_views}
\end{equation}
where $t_a(\cdot)$ and $t_b(\cdot)$ denote two augmentation operators.
In practice, they may correspond to weak and strong augmentations.

To keep the formulation general, we denote by $P$ the number of feature stages used by the victim model.
For single-stage models such as CC, $P=1$; for multi-stage models such as PCLC, $P>1$ \cite{li2021contrastive,zhou2025pclc}.
At stage $p\in\{1,\dots,P\}$, the backbone encoder outputs
\begin{equation}
\begin{aligned}
h_{p,i}^{a} &= f_{\theta}^{p}(x_i^{a}),\\
h_{p,i}^{b} &= f_{\theta}^{p}(x_i^{b}),
\end{aligned}
\label{eq:stage_feature}
\end{equation}
where $f_{\theta}^{p}(\cdot)$ denotes the stage-$p$ feature extractor.
These features are further mapped by an instance-level projector $g_{p}^{I}(\cdot)$ and a cluster-level projector $g_{p}^{C}(\cdot)$:
\begin{equation}
\begin{aligned}
z_{p,i}^{a} &= g_{p}^{I}(h_{p,i}^{a}),\\
z_{p,i}^{b} &= g_{p}^{I}(h_{p,i}^{b}),
\end{aligned}
\label{eq:instance_embedding}
\end{equation}
\begin{equation}
\begin{aligned}
y_{p,i}^{a} &= g_{p}^{C}(h_{p,i}^{a}),\\
y_{p,i}^{b} &= g_{p}^{C}(h_{p,i}^{b}),
\end{aligned}
\label{eq:cluster_prob}
\end{equation}
where $z_{p,i}^{a},z_{p,i}^{b}$ are instance embeddings and $y_{p,i}^{a},y_{p,i}^{b}\in\mathbb{R}^{K}$ are soft cluster-assignment probabilities over $K$ clusters.

\subsubsection{Instance-Level Contrastive Objective}

At stage $p$, the positive pair for sample $x_i$ is $(z_{p,i}^{a}, z_{p,i}^{b})$, while embeddings from different samples in the same mini-batch are treated as negatives, as in InfoNCE/NT-Xent style objectives \cite{oord2018cpc,chen2020simclr,khosla2020supcon}.
We define cosine similarity as
\begin{equation}
\mathrm{sim}(\alpha,\beta)=\frac{\alpha^{\top}\beta}{\|\alpha\|\,\|\beta\|}.
\label{eq:cosine_sim}
\end{equation}

For compact notation, we define the normalization terms
\begin{equation}
\begin{aligned}
\Omega_{p,i}^{a}
&=
\sum_{j=1}^{B}
\exp\!\left(
\frac{\mathrm{sim}(z_{p,i}^{a},z_{p,j}^{b})}{\tau_I}
\right)\\
&\quad+
\sum_{\substack{j=1\\ j\neq i}}^{B}
\exp\!\left(
\frac{\mathrm{sim}(z_{p,i}^{a},z_{p,j}^{a})}{\tau_I}
\right),
\end{aligned}
\label{eq:omega_a}
\end{equation}
\begin{equation}
\begin{aligned}
\Omega_{p,i}^{b}
&=
\sum_{j=1}^{B}
\exp\!\left(
\frac{\mathrm{sim}(z_{p,i}^{b},z_{p,j}^{a})}{\tau_I}
\right)\\
&\quad+
\sum_{\substack{j=1\\ j\neq i}}^{B}
\exp\!\left(
\frac{\mathrm{sim}(z_{p,i}^{b},z_{p,j}^{b})}{\tau_I}
\right),
\end{aligned}
\label{eq:omega_b}
\end{equation}
where $\tau_I$ is the instance-level temperature.

The two directional instance-level losses are
\begin{equation}
\ell_{p,i}^{a}
=
-\log
\frac{
\exp\!\left(\mathrm{sim}(z_{p,i}^{a},z_{p,i}^{b})/\tau_I\right)
}{
\Omega_{p,i}^{a}
},
\label{eq:ins_loss_a}
\end{equation}
\begin{equation}
\ell_{p,i}^{b}
=
-\log
\frac{
\exp\!\left(\mathrm{sim}(z_{p,i}^{b},z_{p,i}^{a})/\tau_I\right)
}{
\Omega_{p,i}^{b}
}.
\label{eq:ins_loss_b}
\end{equation}
Thus, the instance-level loss at stage $p$ is
\begin{equation}
\mathcal{L}_{\mathrm{ins}}^{p}
=
\frac{1}{2B}
\sum_{i=1}^{B}
\left(\ell_{p,i}^{a}+\ell_{p,i}^{b}\right).
\label{eq:instance_loss}
\end{equation}

This objective encourages view-invariant representation learning and determines which features are treated as stable instance-level cues.

\subsubsection{Cluster-Level Contrastive Objective}

Besides instance discrimination, deep contrastive clustering also organizes the learned embeddings into cluster-friendly structures \cite{ji2019iic,li2021contrastive,caron2020swav}.
At stage $p$, we stack the soft assignments of a mini-batch into
\begin{equation}
\begin{aligned}
Y_{p}^{a}
&=
[y_{p,1}^{a},\ldots,y_{p,B}^{a}]^{\top}
\in\mathbb{R}^{B\times K},\\
Y_{p}^{b}
&=
[y_{p,1}^{b},\ldots,y_{p,B}^{b}]^{\top}
\in\mathbb{R}^{B\times K}.
\end{aligned}
\label{eq:cluster_matrix}
\end{equation}

Let $\mathbf{y}_{p,k}^{a}$ and $\mathbf{y}_{p,k}^{b}$ denote the $k$-th columns of $Y_{p}^{a}$ and $Y_{p}^{b}$, respectively.
Each column describes the distribution of mini-batch samples assigned to cluster $k$ under one view.

The positive cluster pair is $(\mathbf{y}_{p,k}^{a}, \mathbf{y}_{p,k}^{b})$, while the remaining columns serve as negatives.
We define the normalization terms
\begin{equation}
\begin{aligned}
\Psi_{p,k}^{a}
&=
\sum_{j=1}^{K}
\exp\!\left(
\frac{\mathrm{sim}(\mathbf{y}_{p,k}^{a},\mathbf{y}_{p,j}^{b})}{\tau_C}
\right)\\
&\quad+
\sum_{\substack{j=1\\ j\neq k}}^{K}
\exp\!\left(
\frac{\mathrm{sim}(\mathbf{y}_{p,k}^{a},\mathbf{y}_{p,j}^{a})}{\tau_C}
\right),
\end{aligned}
\label{eq:psi_a}
\end{equation}
\begin{equation}
\begin{aligned}
\Psi_{p,k}^{b}
&=
\sum_{j=1}^{K}
\exp\!\left(
\frac{\mathrm{sim}(\mathbf{y}_{p,k}^{b},\mathbf{y}_{p,j}^{a})}{\tau_C}
\right)\\
&\quad+
\sum_{\substack{j=1\\ j\neq k}}^{K}
\exp\!\left(
\frac{\mathrm{sim}(\mathbf{y}_{p,k}^{b},\mathbf{y}_{p,j}^{b})}{\tau_C}
\right),
\end{aligned}
\label{eq:psi_b}
\end{equation}
where $\tau_C$ is the cluster-level temperature.

The two directional cluster-level losses are
\begin{equation}
\hat{\ell}_{p,k}^{a}
=
-\log
\frac{
\exp\!\left(\mathrm{sim}(\mathbf{y}_{p,k}^{a},\mathbf{y}_{p,k}^{b})/\tau_C\right)
}{
\Psi_{p,k}^{a}
},
\label{eq:clu_loss_a}
\end{equation}
\begin{equation}
\hat{\ell}_{p,k}^{b}
=
-\log
\frac{
\exp\!\left(\mathrm{sim}(\mathbf{y}_{p,k}^{b},\mathbf{y}_{p,k}^{a})/\tau_C\right)
}{
\Psi_{p,k}^{b}
}.
\label{eq:clu_loss_b}
\end{equation}

To prevent trivial collapse, we regularize the average cluster distributions:
\begin{equation}
\begin{aligned}
\bar{y}_{p}^{a}
&=
\frac{1}{B}\sum_{i=1}^{B}y_{p,i}^{a},\\
\bar{y}_{p}^{b}
&=
\frac{1}{B}\sum_{i=1}^{B}y_{p,i}^{b},
\end{aligned}
\label{eq:avg_prob}
\end{equation}
\begin{equation}
\mathcal{H}(\bar{y})
=
-\sum_{k=1}^{K}\bar{y}_{k}\log \bar{y}_{k}.
\label{eq:entropy}
\end{equation}
The final cluster-level loss at stage $p$ is
\begin{equation}
\begin{aligned}
\mathcal{L}_{\mathrm{clu}}^{p}
&=
\frac{1}{2K}
\sum_{k=1}^{K}
\left(\hat{\ell}_{p,k}^{a}+\hat{\ell}_{p,k}^{b}\right)\\
&\quad
-\mathcal{H}(\bar{y}_{p}^{a})
-\mathcal{H}(\bar{y}_{p}^{b}).
\end{aligned}
\label{eq:cluster_loss}
\end{equation}

This objective governs how local embeddings are aggregated into a global cluster structure.
Therefore, once trigger-biased representations start to emerge, cluster-level consistency can further consolidate and propagate the bias.

\subsubsection{Overall Objective of the Victim Model}

The complete objective of the victim deep contrastive clustering model is
\begin{equation}
\begin{aligned}
\mathcal{L}_{\mathrm{all}}
&=
\sum_{p=1}^{P}
\left(
\mathcal{L}_{\mathrm{ins}}^{p}
+
\mathcal{L}_{\mathrm{clu}}^{p}
\right).
\end{aligned}
\label{eq:overall_loss}
\end{equation}

This dual-level design is central to our attack.
A persistent trigger can first dominate instance-level invariance learning and then bias cluster-level assignment consistency.
As a consequence, the backdoor is not encoded as a simple input-output shortcut, but as a structural bias in the learned clustering space.

\subsection{Trigger-Based Poisoning Strategy}
\label{sec:poisoning_strategy}
\begin{figure}[t]
    \centering
    \includegraphics[width=1\columnwidth]{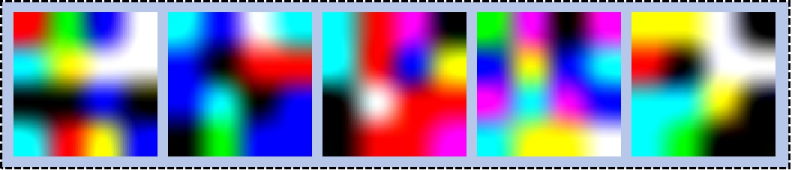}
    \caption{Examples of trigger patterns in the trigger bank. Different poisoned subsets can be associated with different trigger patterns, enabling both single-trigger and multi-trigger poisoning settings.}
    \label{fig:trigger_bank}
\end{figure}

Let $\mathcal{T}_{\delta}=\{\delta_m\}_{m=1}^{M}$ denote a trigger bank with $M$ trigger patterns, following the common trigger-based backdoor formulation \cite{gu2019badnets,saha2020hidden,nguyen2021wanet}. Examples of different trigger patterns are shown in Fig.~\ref{fig:trigger_bank}.

The trigger bank is selected according to four practical criteria.
First, the trigger should be localized and should not dominate the image semantics, so that poisoned images remain usable training samples.
Second, it should remain stable under common contrastive augmentations, such as random cropping, resizing, flipping, and color perturbation.
Third, it should be learnable by the victim encoder, so that the model can consistently encode it as a representation cue during training.
Fourth, the bank should contain diverse patterns, allowing different poisoned subsets to form different trigger-associated regions rather than collapsing into a single trigger type.

The attacker samples a poisoned subset $\mathcal{S}\subset\mathcal{D}$ with poisoning ratio
\begin{equation}
\rho=\frac{|\mathcal{S}|}{|\mathcal{D}|}.
\label{eq:poison_ratio}
\end{equation}
In the more general multi-trigger setting, the poisoned subset can be decomposed as
\begin{equation}
\mathcal{S}
=
\biguplus_{m=1}^{M}\mathcal{S}_{m},
\label{eq:poison_subset}
\end{equation}
where the subsets are disjoint and each subset $\mathcal{S}_{m}$ is associated with one trigger pattern $\delta_m$.

For each clean image $x\in\mathcal{S}_{m}$, the attacker injects the trigger by
\begin{equation}
\tilde{x}
=
\mathcal{P}(x,\delta_m,r,\alpha),
\label{eq:trigger_injection}
\end{equation}
where $r$ is a randomized spatial position and $\alpha$ controls the trigger strength or blending ratio \cite{chen2017targeted,saha2020hidden,liao2018invisible}.
The resulting poisoned dataset is
\begin{equation}
\tilde{\mathcal{D}}
=
(\mathcal{D}\setminus\mathcal{S})
\cup
\left\{
\mathcal{P}(x,\delta_m,r,\alpha)
\,\middle|\,
x\in\mathcal{S}_m,\; m=1,\dots,M
\right\}.
\label{eq:poisoned_dataset}
\end{equation}

This design satisfies three practical requirements.
First, the poisoning budget is small, which helps preserve normal clean-data behavior.
Second, randomized trigger placement avoids overly regular artifacts and prevents the attack from relying on a fixed position.
Third, the trigger remains limited in size and strength while still being stable enough to survive common augmentations, which is crucial for successful implantation under contrastive training \cite{saha2022sslbackdoor,zhang2024corruptencoder,zhang2025invisible}.

\subsection{Backdoor Implantation under Dual-Level Objectives}
\label{sec:implantation}
The victim model is trained on $\tilde{\mathcal{D}}$ using the original dual-level objective:
\begin{equation}
\begin{aligned}
\tilde{\mathcal{L}}_{\mathrm{all}}
&=
\sum_{p=1}^{P}
\left(
\tilde{\mathcal{L}}_{\mathrm{ins}}^{p}
+
\tilde{\mathcal{L}}_{\mathrm{clu}}^{p}
\right).
\end{aligned}
\label{eq:poisoned_overall_loss}
\end{equation}
The critical point is that the trigger is amplified at both levels.

\noindent\textbf{Instance-level amplification.}
For a poisoned sample $\tilde{x}$, the two views
\begin{equation}
\begin{aligned}
\tilde{x}^{a} &= t_a(\tilde{x}),\\
\tilde{x}^{b} &= t_b(\tilde{x})
\end{aligned}
\label{eq:poisoned_views}
\end{equation}
retain the same trigger cue when the trigger survives the stochastic augmentations, matching the augmentation-invariance mechanism exploited in prior SSL backdoor attacks \cite{saha2022sslbackdoor,zhang2024corruptencoder,chen2025noisyalignment}.
Thus, minimizing $\tilde{\mathcal{L}}_{\mathrm{ins}}^{p}$ encourages the model to encode the trigger as a stable shortcut feature.
Since the trigger is often easier to match than high-level semantics, it becomes a convenient anchor for view alignment.

\noindent\textbf{Cluster-level reinforcement.}
Once multiple poisoned samples exhibit similar trigger-dominated embeddings, they also tend to produce similar soft cluster assignments.
Minimizing $\tilde{\mathcal{L}}_{\mathrm{clu}}^{p}$ then further consolidates these biased assignments, thereby stabilizing the trigger-associated cluster region.

Together, these two effects produce a \emph{dual-level amplification mechanism}:
\begin{itemize}
    \item \textbf{Instance-level attraction:} triggered samples are pulled together in representation space because they share a common easy-to-align pattern;
    \item \textbf{Cluster-level reinforcement:} consistent soft assignments propagate this local bias into the global cluster structure.
\end{itemize}

This coupled optimization process gradually transforms a local persistent cue into a malicious clustering prior.
Therefore, HBDCC attacks not a single component of the victim model, but the interaction between view-invariant representation learning and cluster-consistent grouping.

\subsection{Inference-Time Activation and Cluster Distortion}
\label{sec:activation}

After training, the poisoned model usually behaves normally on clean inputs because the trigger-dependent region is not activated.
At inference time, once a trigger is inserted into a test image,
\begin{equation}
x^{*}
=
\mathcal{P}(x,\delta_m,r,\alpha),
\label{eq:test_trigger}
\end{equation}
its representation is drawn toward the trigger-associated region learned during training.
As a result, the final cluster assignment can be manipulated regardless of the image's original semantic identity.

This induces two typical forms of structural corruption:
\begin{itemize}
    \item \textbf{Cross-class confusion:} semantically unrelated samples are grouped into the same cluster because they share similar trigger-induced representations;
    \item \textbf{Class splitting:} samples that should belong to one semantic class are divided into multiple clusters after different triggers are injected.
\end{itemize}

Let $\mathcal{C}_{k}$ denote the set of samples assigned to cluster $k$, and let $z_i$ denote the final representation used for clustering.
The centroid of cluster $k$ is
\begin{equation}
c_k
=
\frac{1}{|\mathcal{C}_k|}
\sum_{z_i\in\mathcal{C}_k} z_i.
\label{eq:centroid}
\end{equation}
As poisoned samples accumulate, the centroids or prototypes are progressively shifted toward trigger-dominated regions.
Because this process is gradual and distributed across the representation space, it is not reliably exposed by clean-set clustering evaluation alone.
The backdoor thus manifests as \emph{semantic disorganization of the cluster structure}, rather than direct label flipping as in supervised attacks.

\subsection{Knowledge-Distillation-Based Mitigation}
\label{sec:mitigation}

To mitigate trigger dependence, we further employ a knowledge-distillation-based defense \cite{bucilua2006model,hinton2015distilling,romero2015fitnets}.
The idea is to use a small clean unlabeled subset to transfer benign clustering knowledge from a clean teacher model to a student model, thereby weakening the malicious association between triggers and cluster semantics.
In this defense, the teacher is a clean model trained without poisoned samples, while the student is initialized from the poisoned model that needs repair.

Let $F_T$ denote the clean teacher model and $F_S$ the student model.
Given a clean unlabeled subset $\mathcal{D}_c$, the student is optimized by
\begin{equation}
\begin{aligned}
\mathcal{L}_{\mathrm{KD}}
&=
\lambda_1 \mathcal{L}_{\mathrm{ins}}^{S}
+ \lambda_2 \mathcal{L}_{\mathrm{clu}}^{S}\\
&\quad
+ \lambda_3 \mathcal{L}_{\mathrm{distill}},
\end{aligned}
\label{eq:kd_loss}
\end{equation}
where $\mathcal{L}_{\mathrm{distill}}$ encourages the student to match the clean clustering behavior of the teacher, and $\lambda_1,\lambda_2,\lambda_3$ are balancing coefficients.
Let $q_{T,i}$ and $q_{S,i}$ denote the teacher and student soft cluster distributions for the $i$-th clean sample in a mini-batch.
We instantiate the distillation term as
\begin{equation}
\mathcal{L}_{\mathrm{distill}}
=
\frac{1}{B}
\sum_{i=1}^{B}
D_{\mathrm{KL}}
\left(
q_{T,i}\,\|\,q_{S,i}
\right),
\label{eq:distill_loss}
\end{equation}
where $D_{\mathrm{KL}}(\cdot\|\cdot)$ denotes the Kullback--Leibler divergence.

Because the student is distilled only on clean unlabeled data, the trigger-related correlation is no longer reinforced.
Instead, the model inherits the benign semantic structure while suppressing backdoor-dependent behavior.
This defense does not assume label access and is therefore compatible with the unsupervised setting considered in this paper.

\subsection{Overall Attack Algorithm}
\label{sec:algorithm}

Algorithm~\ref{alg:hbdcc} summarizes the overall pipeline of HBDCC.

\begin{algorithm}[t]
\caption{HBDCC Attack: Hidden Backdoor for Deep Contrastive Clustering}
\label{alg:hbdcc}
\begin{algorithmic}[1]
\Require Unlabeled dataset $\mathcal{D}$, trigger bank $\mathcal{T}_{\delta}$, poisoning ratio $\rho$, number of stages $P$, victim model $F_{\theta}$
\Ensure Poisoned model $F_{\theta}^{*}$
\State Sample poisoned subsets $\mathcal{S}_{1},\dots,\mathcal{S}_{M}$ from $\mathcal{D}$ according to $\rho$
\For{each subset $\mathcal{S}_{m}$}
    \For{each sample $x \in \mathcal{S}_{m}$}
        \State Randomly sample trigger location $r$
        \State Generate poisoned sample $\tilde{x}=\mathcal{P}(x,\delta_m,r,\alpha)$
    \EndFor
\EndFor
\State Construct poisoned dataset $\tilde{\mathcal{D}}=(\mathcal{D}\setminus\mathcal{S})\cup\{\tilde{x}\}$
\For{epoch $=1,2,\dots,E$}
    \For{mini-batch $\mathcal{B}\subset\tilde{\mathcal{D}}$}
        \State Generate two augmented views $\mathcal{B}^{a}$ and $\mathcal{B}^{b}$
        \For{$p=1$ to $P$}
            \State Compute stage-$p$ features $h_{p}^{a}, h_{p}^{b}$
            \State Compute instance embeddings $z_{p}^{a}, z_{p}^{b}$ using $g_{p}^{I}$
            \State Compute cluster assignments $y_{p}^{a}, y_{p}^{b}$ using $g_{p}^{C}$
            \State Compute $\mathcal{L}_{\mathrm{ins}}^{p}$ using Eq.~\eqref{eq:instance_loss}
            \State Compute $\mathcal{L}_{\mathrm{clu}}^{p}$ using Eq.~\eqref{eq:cluster_loss}
        \EndFor
        \State Compute $\mathcal{L}_{\mathrm{all}}$ using Eq.~\eqref{eq:overall_loss}
        \State Update $\theta$ by gradient descent on $\mathcal{L}_{\mathrm{all}}$
\EndFor
\EndFor
\State Obtain poisoned model $F_{\theta}^{*}$
\For{each test sample $x$}
    \State Inject a trigger and obtain $x^{*}=\mathcal{P}(x,\delta_m,r,\alpha)$
    \State Feed $x^{*}$ into $F_{\theta}^{*}$ and output its cluster assignment
\EndFor
\end{algorithmic}
\end{algorithm}
Overall, HBDCC is a pure data-poisoning attack that requires neither label access nor any modification to the victim training pipeline.
Its effectiveness stems from a structural vulnerability of deep contrastive clustering: the coupled optimization of instance-level invariance and cluster-level consistency unintentionally provides a natural channel for backdoor implantation.

\section{Experiments} 
\label{sec:experiments}

In this section, we evaluate the proposed \textbf{HBDCC} on three benchmark image clustering datasets and analyze attack effectiveness, clean-data concealment, cross-model behavior, trigger sensitivity, and mitigation, following standard image-clustering evaluation practice \cite{xie2016dec,caron2018deepcluster,li2021contrastive}.

We first describe the experimental setup in Section~\ref{sec:exp_setup}.
We then study the attack behavior under different poisoning ratios in Section~\ref{sec:poison_ratio}, followed by a comparison between single-class and multi-class poisoning in Section~\ref{sec:single_multi}.
Next, we validate the attack on different contrastive clustering architectures in Section~\ref{sec:cross_model}, and further analyze the effects of trigger type and trigger size in Sections~\ref{sec:trigger_type} and~\ref{sec:trigger_size}, respectively.
We finally evaluate a knowledge-distillation-based defense in Section~\ref{sec:defense}.

\subsection{Experimental Setup}
\label{sec:exp_setup}
\begin{figure*}[!t]
	\centering
	\subfigure[ImageNet-10]{
		\includegraphics[width=0.3\textwidth]{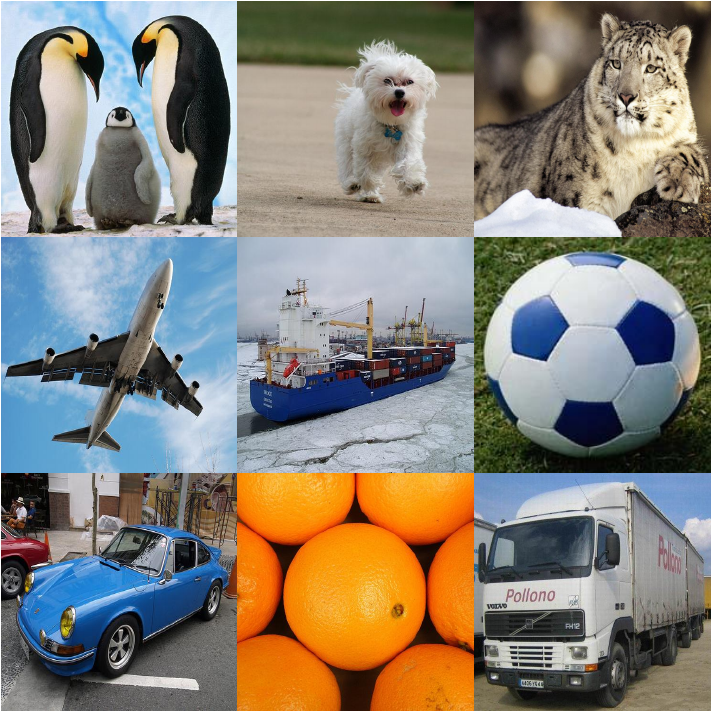}}
	\subfigure[ImageNet-Dogs]{
		\includegraphics[width=0.3\textwidth]{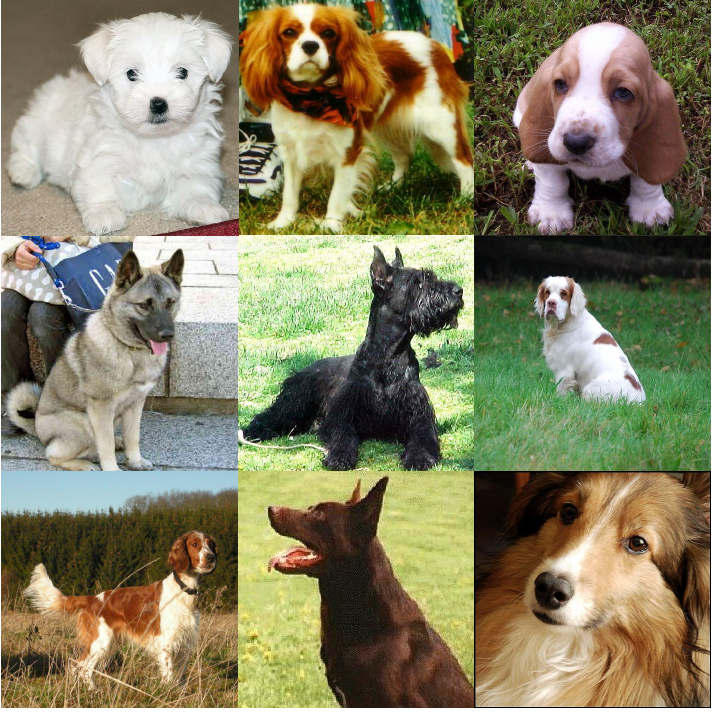}}
	\subfigure[STL-10]{
		\includegraphics[width=0.3\textwidth]{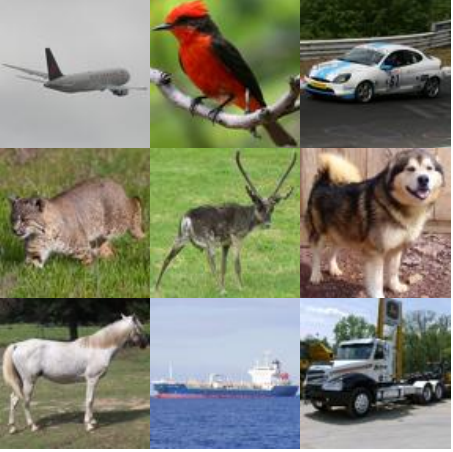}}
	\caption{Representative examples of the three real-world image datasets used in our experiments, including ImageNet-10, ImageNet-Dogs, and STL-10.}
	\label{fig:dataset_samples}
\end{figure*}

\noindent\textbf{Datasets.}
We conduct experiments on three widely used image clustering benchmarks: \textbf{ImageNet-10}, \textbf{ImageNet-Dogs}, and \textbf{STL-10} \cite{deng2009imagenet,coates2011stl,li2021contrastive}.
Representative examples are shown in Fig.~\ref{fig:dataset_samples}.
ImageNet-10 contains 13,000 images from 10 semantic categories.
ImageNet-Dogs is a fine-grained benchmark constructed from ImageNet, containing 19,500 images from 15 dog breeds \cite{deng2009imagenet}.
STL-10 contains 13,000 images from 10 categories \cite{coates2011stl}.
These three datasets provide diverse evaluation scenarios, ranging from relatively separable object categories to more challenging fine-grained clustering settings.

\begin{table}[t]
    \centering
    \caption{Datasets used in our experiments.}
    \label{tab:datasets}
    \renewcommand{\arraystretch}{1.1}H
    \begin{tabular}{C{0.42\columnwidth}C{0.24\columnwidth}C{0.17\columnwidth}}
        \toprule
        \midrule
        Dataset & \#Samples & \#Classes \\
        \midrule
        ImageNet-10   & 13,000 & 10 \\
        ImageNet-Dogs & 19,500 & 15 \\
        STL-10        & 13,000 & 10 \\
        \bottomrule
    \end{tabular}
\end{table}

\noindent\textbf{Victim model and training protocol.}
We use \textbf{Contrastive Clustering (CC)} as the primary victim model.
CC is a representative end-to-end contrastive clustering framework that contains both an instance-level projector and a cluster-level projector, making it a suitable target for backdoor implantation \cite{li2021contrastive}.
Unless otherwise specified, we train CC from scratch for 1,000 epochs with a batch size of 256.
All experiments follow the official training configuration of CC to ensure reproducibility and fair comparison across clean and poisoned runs.
The main CC experiments are conducted on a single NVIDIA RTX 3090 GPU with CUDA 11.0.

\noindent\textbf{Trigger construction.}
We adopt triggers based on Hidden Trigger Backdoor Attacks
(HTBA) \cite{saha2020hidden}.
Specifically, each trigger is generated from a randomly initialized $4 \times 4$ RGB image and resized to the desired patch size by bilinear interpolation.
Compared with supervised classification, clustering is intrinsically more difficult to attack because multiple semantic categories coexist in the same representation space.
Therefore, we further adjust the trigger shape, size, and spatial position to make the trigger more suitable for contrastive clustering scenarios.
For multi-class poisoning, each attacked class is associated with a distinct trigger during training, and the trigger is randomly placed on the image to avoid a fixed-position shortcut.
The use of class identities in this controlled experiment is only for defining source subsets and evaluating attack scope; the HBDCC pipeline itself does not require label supervision during victim training.

\noindent\textbf{Evaluation metrics.}
We follow the standard clustering protocol and report three widely used metrics: \textbf{normalized mutual information (NMI)}, \textbf{adjusted Rand index (ARI)}, and \textbf{clustering accuracy (ACC)} \cite{strehl2002cluster,vinh2010information,hubert1985comparing,munkres1957algorithms}.
Higher values indicate better clustering performance.
In our setting, a successful backdoor attack should satisfy two conditions simultaneously:
(1) the poisoned model should remain close to the clean model on clean data, and
(2) the poisoned model should suffer a substantial performance drop on triggered inputs.

\subsection{Attack Performance under Different Poisoning Ratios}
\label{sec:poison_ratio}
We first study the effectiveness of HBDCC under different poisoning ratios.
For each dataset, we construct poisoned training sets with poisoning rates of 5\%, 10\%, 15\%, and 20\%.
In the multi-class setting, each class is assigned a class-specific trigger during training.
At test time, triggered samples are generated by inserting \emph{mismatched} triggers into the corresponding images, thereby activating the backdoor and forcing the model to produce incorrect cluster assignments.

Before training poisoned models, we first check whether the trigger alone is harmful.
Table~\ref{tab:clean_model_triggered_test} shows that directly applying triggers to a clean model causes only minor changes.
On ImageNet-10, for example, ACC decreases by only 0.5 points, from 89.5 to 89.0; on STL-10, the corresponding drop is 0.3 points.
The same pattern holds on ImageNet-Dogs.
Thus, the trigger pattern is not simply an input-level corruption that breaks clustering by itself.
The degradation observed later must come from the model learning a trigger-dependent association during poisoned training.
\begin{table}[t]
    \centering
    \caption{Performance of the clean model on clean and triggered test sets.}
    \label{tab:clean_model_triggered_test}
    \setlength{\tabcolsep}{3.5pt}
    \renewcommand{\arraystretch}{1.15}
    \begin{tabular}{C{0.27\columnwidth}|*{3}{C{0.085\columnwidth}}|*{3}{C{0.085\columnwidth}}}
        \toprule
        \midrule
        \multirow{2}{*}{Dataset} & \multicolumn{3}{c|}{\rule{0pt}{2.4ex}Clean inputs} & \multicolumn{3}{c}{\rule{0pt}{2.4ex}Triggered inputs} \\
        \cline{2-7}
        & \rule{0pt}{2.4ex}NMI & ARI & ACC & NMI & ARI & ACC \\
        \midrule
        ImageNet-10   & 86.2 & 82.5 & 89.5 & 85.6 & 81.8 & 89.0 \\
        ImageNet-Dogs & 41.9 & 24.0 & 36.6 & 41.5 & 23.0 & 36.5 \\
        STL-10        & 67.5 & 60.8 & 74.8 & 66.9 & 59.6 & 74.5 \\
        \bottomrule
    \end{tabular}
\end{table}

We then train poisoned CC models using datasets with different poisoning ratios.
Table~\ref{tab:poisoned_model_ratio} reports the results of the \emph{poisoned model} on clean and triggered test sets.
Several important observations can be made.

Three trends are evident.
First, clean-data utility is mostly preserved.
On ImageNet-10, clean ACC stays in a narrow range from 88.4 to 90.2 across all poisoning ratios, and on STL-10 it remains between 74.5 and 78.7.
ImageNet-Dogs is less stable because fine-grained dog breeds are harder to separate; the 10\% setting, for instance, shows a clear ACC drop on clean inputs.
Even in this harder case, however, NMI and ARI remain close to the clean baseline in most poisoned runs, and the degradation is much smaller than that observed under triggered evaluation.
This behavior is important for the hidden nature of the attack: standard clean-set evaluation alone would not reliably reveal the implanted backdoor.

Second, triggered inputs expose a large hidden failure mode.
At 10\% poisoning, ImageNet-10 drops from 88.4 ACC on clean inputs to 73.3 ACC on triggered inputs.
The effect is much stronger on the two harder settings: ImageNet-Dogs falls to 10.1 ACC, and STL-10 falls to 10.4 ACC.
At 20\% poisoning, ImageNet-Dogs and STL-10 are close to random-level assignment accuracy, while ImageNet-10 still loses more than 30 ACC points.
The attack therefore does not merely add noise to the output; it changes the organization of the learned representation whenever the trigger is present.

\begin{figure*}[!t]
	\centering
	\subfigure[ImageNet-10 (Before)]{
		\includegraphics[width=0.31\textwidth]{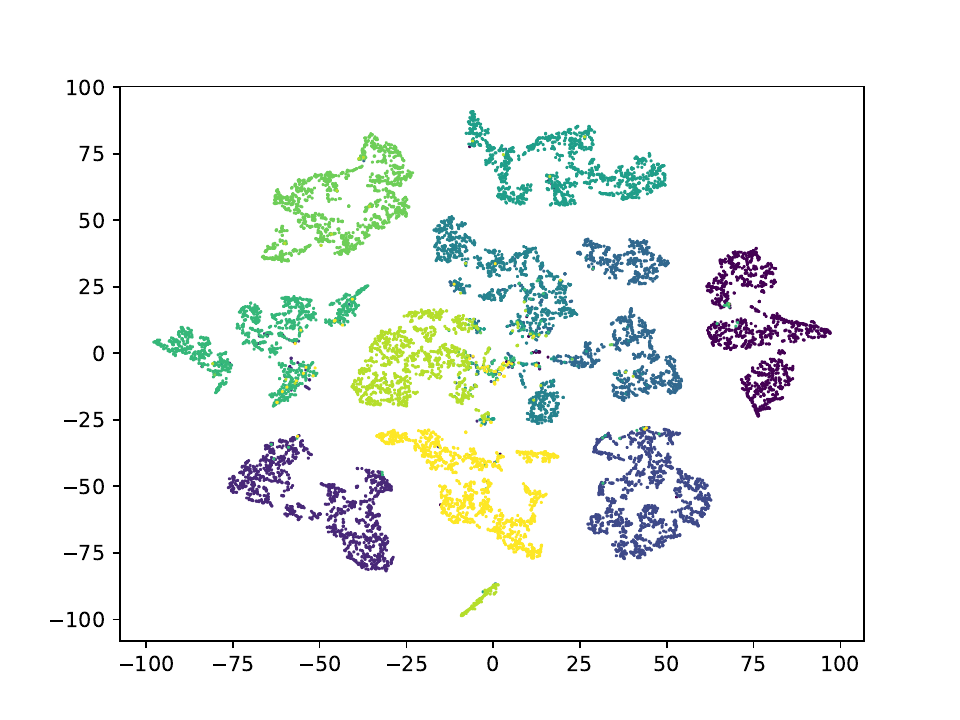}}
	\subfigure[ImageNet-Dogs (Before)]{
		\includegraphics[width=0.31\textwidth]{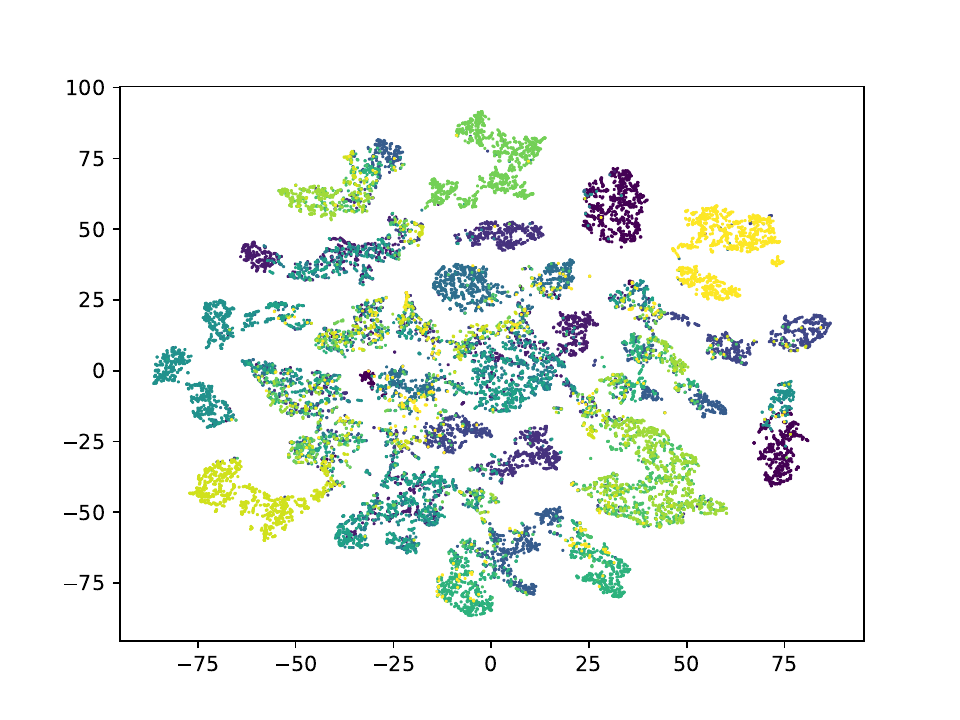}}
	\subfigure[STL-10 (Before)]{
		\includegraphics[width=0.31\textwidth]{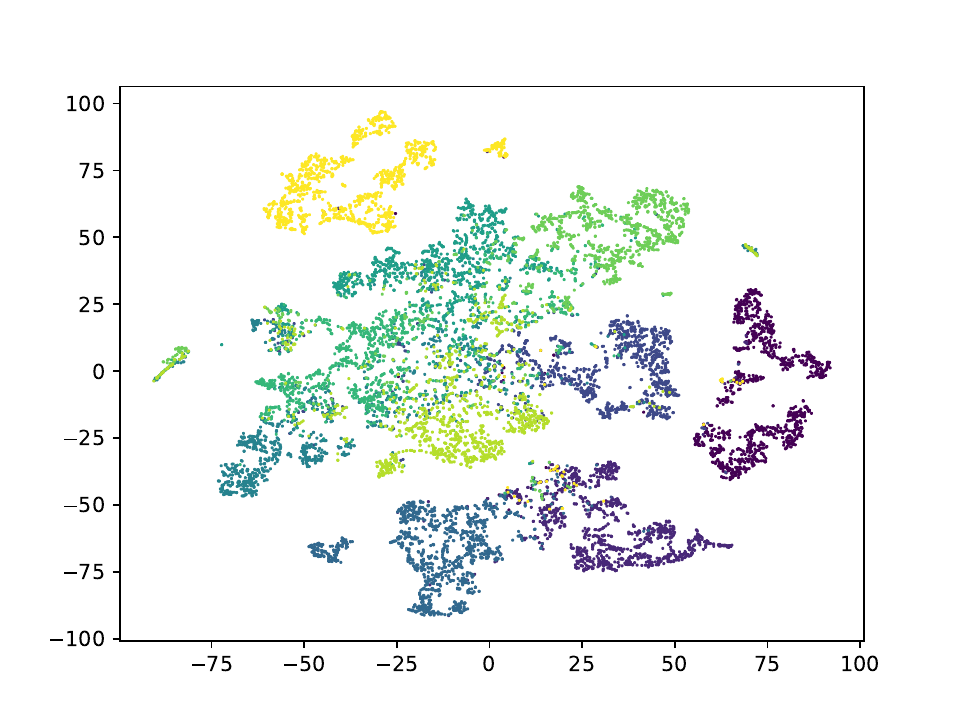}}
	\vspace{0.5em}
	\subfigure[ImageNet-10 (After)]{
		\includegraphics[width=0.31\textwidth]{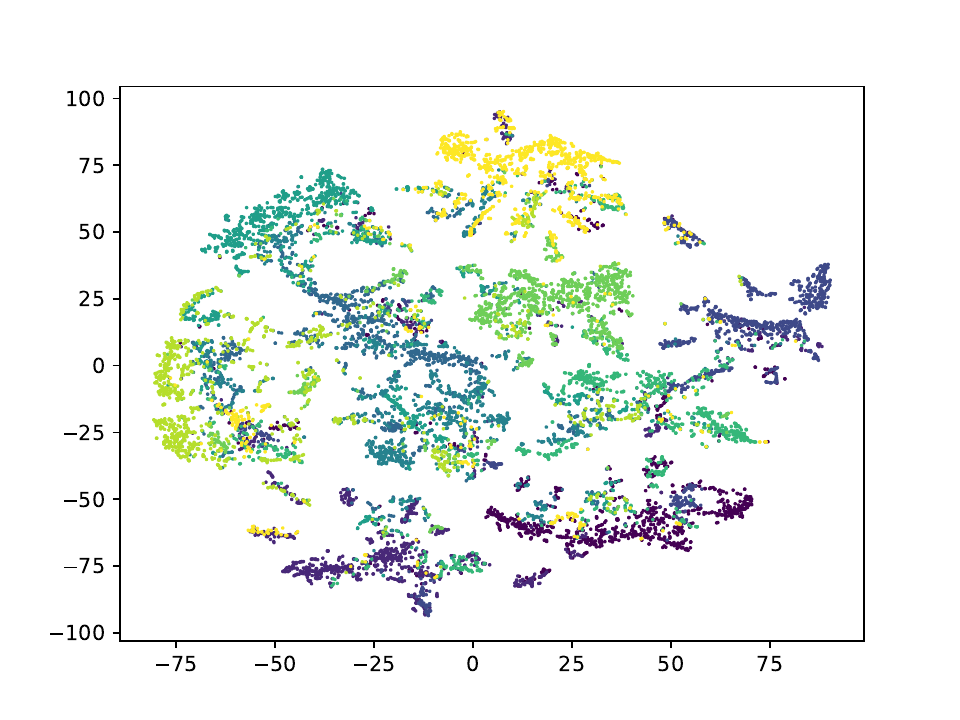}}
	\subfigure[ImageNet-Dogs (After)]{
		\includegraphics[width=0.31\textwidth]{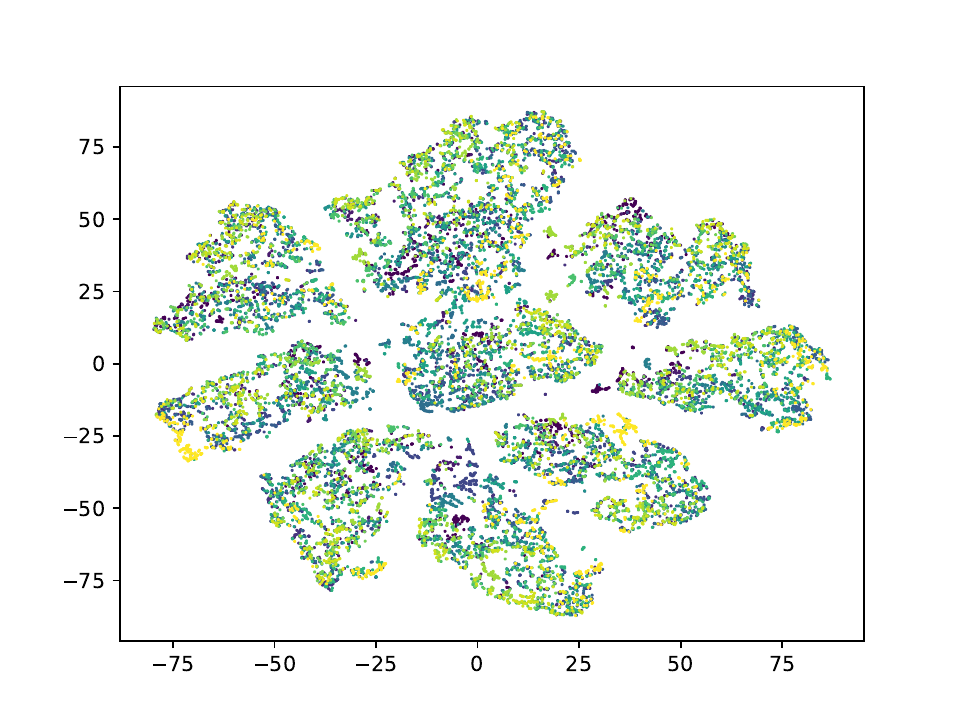}}
	\subfigure[STL-10 (After)]{
		\includegraphics[width=0.31\textwidth]{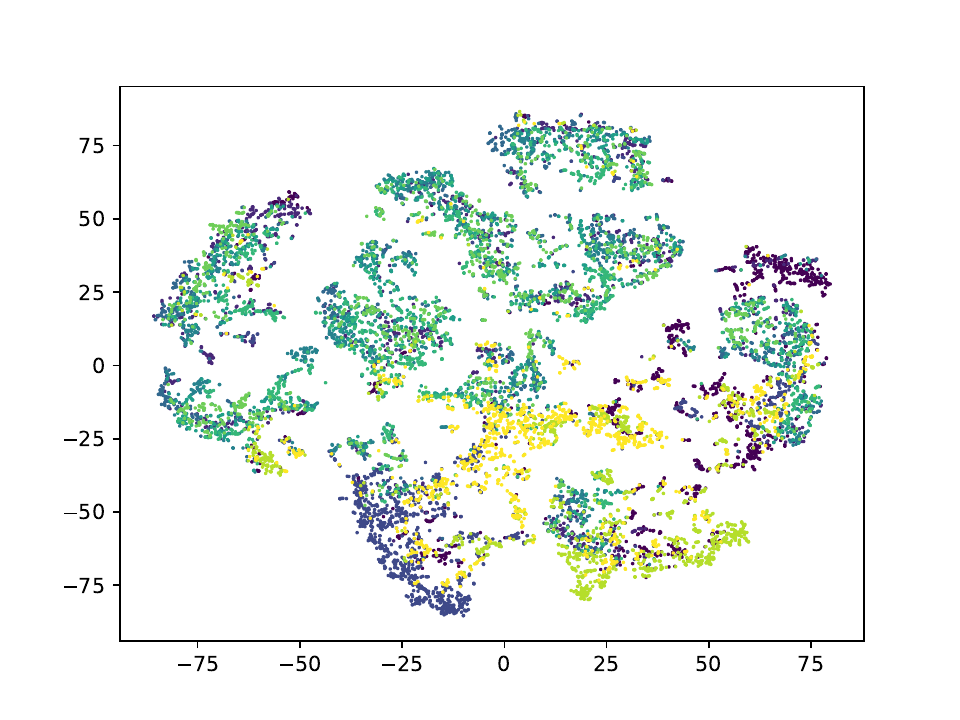}}
	\caption{t-distributed stochastic neighbor embedding (t-SNE) visualizations \cite{maaten2008tsne} of the learned representations on ImageNet-10, ImageNet-Dogs, and STL-10 before and after the backdoor attack. Before the attack, the clean models produce relatively compact and separable clusters. After the attack, the cluster structure becomes significantly distorted, with more severe cross-class mixing and degraded semantic consistency.}
	\label{fig:tsne_attack_all}

\end{figure*}

Third, the three metrics reveal different aspects of the failure.
ACC and ARI are especially sensitive to cluster assignment corruption, while NMI can remain comparatively higher when the model still preserves some coarse grouping structure.
This explains the non-monotonic NMI on STL-10 at 10\% poisoning: the triggered samples retain partial mutual-information structure, but their label-to-cluster alignment is already severely damaged.
The t-SNE visualization in Fig.~\ref{fig:tsne_attack_all} is consistent with this interpretation, showing that HBDCC disrupts the geometry of the embedding space rather than only changing final cluster labels.
\begin{table*}[t]
    \centering
    \caption{Clean and triggered performance of poisoned models under different poisoning ratios.}
    \label{tab:poisoned_model_ratio}
    \small
    \setlength{\tabcolsep}{4.5pt}
    \renewcommand{\arraystretch}{1.18}
    \begin{tabular}{C{0.12\textwidth}|C{0.16\textwidth}|*{3}{C{0.093\textwidth}}|*{3}{C{0.093\textwidth}}}
        \toprule
        \midrule
        \multirow{2}{*}{\shortstack{Poisoning\\Ratio}} &
        \multirow{2}{*}{Dataset} &
        \multicolumn{3}{c|}{\rule{0pt}{2.4ex}Clean inputs} &
        \multicolumn{3}{c}{\rule{0pt}{2.4ex}Triggered inputs} \\
        \cline{3-8}
        & & \rule{0pt}{2.4ex}NMI & ARI & ACC & NMI & ARI & ACC \\
        \midrule
        \multirow{3}{*}{5\%}
          & ImageNet-10   & 85.7 & 81.9 & 88.9 & 78.7 & 75.3 & 86.8 \\
          & ImageNet-Dogs & 42.3 & 25.3 & 40.5 & 10.1 & 10.0 & 16.7 \\
          & STL-10        & 70.7 & 63.6 & 78.7 & 35.1 &  5.6 &  9.3 \\
        \midrule
        \multirow{3}{*}{10\%}
          & ImageNet-10   & 85.9 & 82.1 & 88.4 & 55.7 & 51.7 & 73.3 \\
          & ImageNet-Dogs & 40.6 & 23.1 & 25.1 &  6.4 & 2.1 & 10.1 \\
          & STL-10        & 68.5 & 61.6 & 76.5 & 43.6 & 17.6 & 10.4 \\
        \midrule
        \multirow{3}{*}{15\%}
          & ImageNet-10   & 84.8 & 81.4 & 88.8 & 40.6 & 36.2 & 61.8 \\
          & ImageNet-Dogs & 41.2 & 25.6 & 37.2 &  1.6 &  0.6 &  9.6 \\
          & STL-10        & 67.3 & 59.3 & 75.1 & 32.3 & 12.9 &  9.6 \\
        \midrule
        \multirow{3}{*}{20\%}
          & ImageNet-10   & 87.2 & 86.2 & 90.2 & 36.8 & 32.6 & 58.4 \\
          & ImageNet-Dogs & 41.4 & 24.7 & 35.1 &  0.6 &  2.1 &  7.2 \\
          & STL-10        & 68.5 & 61.6 & 74.5 & 14.4 & 11.2 &  8.4 \\
        \bottomrule
    \end{tabular}
\end{table*}

\begin{figure}[t]
    \centering
    \includegraphics[width=1\columnwidth]{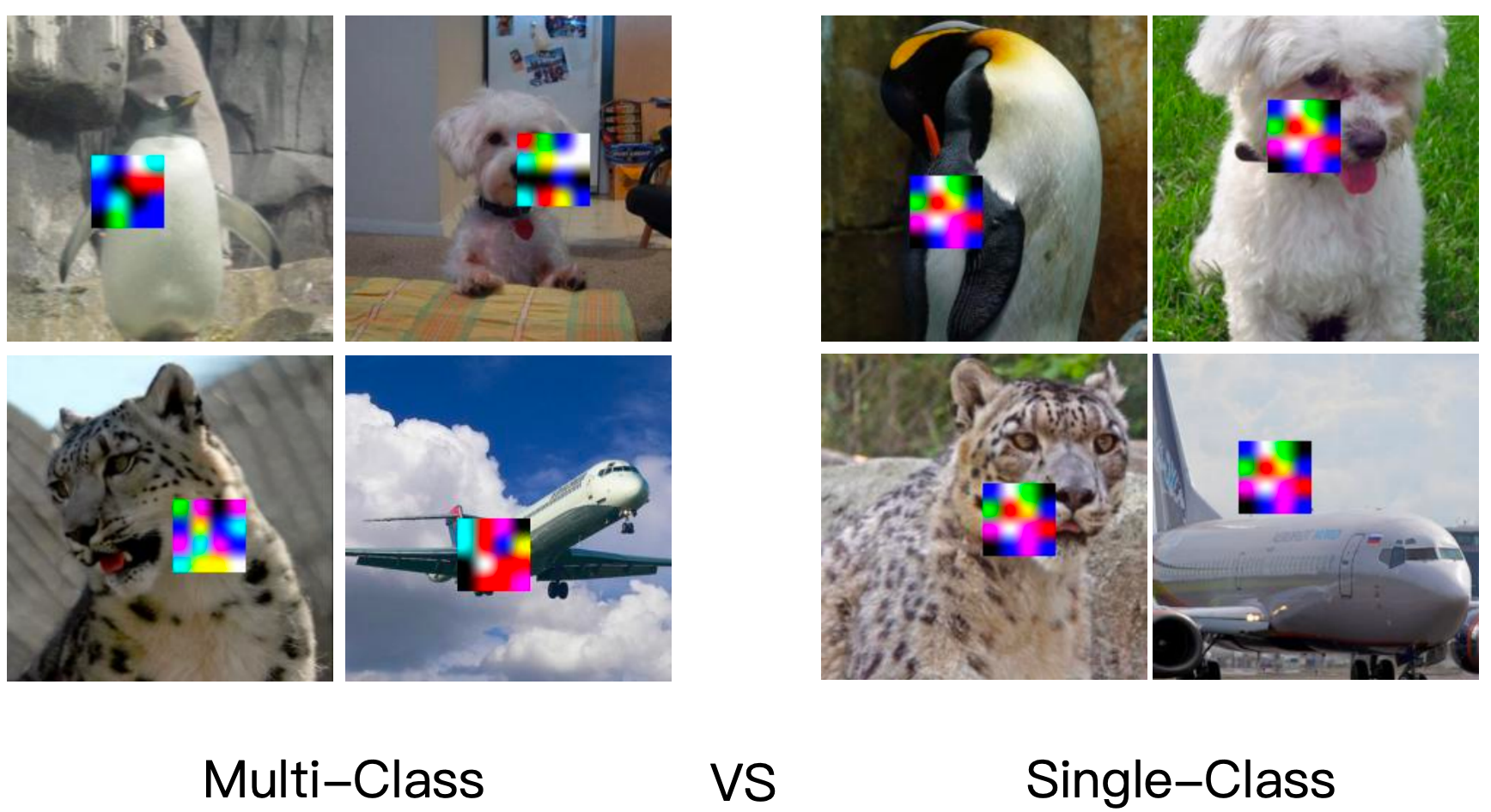}
    \caption{Examples of poisoned samples used in the attack-scope study. In the multi-class setting, different semantic classes can be associated with different triggers; in the single-class setting, one learned trigger association is activated on samples from other classes at test time.}
    \label{fig:attack_scope_examples}
\end{figure}

\subsection{Single-Class versus Multi-Class Poisoning}
\label{sec:single_multi}
The scope of poisoning can substantially affect how a backdoor disrupts the learned cluster structure.
Depending on how trigger-associated samples are distributed across semantic categories, the attack may induce cross-class confusion or fragment the original clustering structure.
We therefore investigate whether poisoning only a single semantic class is sufficient to affect the overall clustering system, and compare it with a broader multi-class poisoning strategy.

In the \emph{single-class poisoning} setting, only one class in ImageNet-10 is poisoned during training, while samples from all other classes remain clean.
At test time, the learned trigger is inserted into samples from other classes to examine whether a trigger association learned from a limited source subset can generalize beyond that subset.
In contrast, under \emph{multi-class poisoning}, multiple semantic classes are poisoned during training with class-specific triggers, producing several trigger-associated regions in the learned representation space.
Representative poisoned samples from the two settings are shown in Fig.~\ref{fig:attack_scope_examples}.

The results are reported in Table~\ref{tab:single_vs_multi}.
Both poisoning strategies preserve relatively strong clustering performance on clean inputs.
Under multi-class poisoning, the poisoned model achieves 85.9/82.1/88.4 in NMI/ARI/ACC on the clean test set, while the corresponding performance under single-class poisoning is 85.7/82.1/89.3.
This indicates that neither attack causes an obvious degradation under standard clean-data evaluation.

A clear difference emerges once the trigger is activated.
Under single-class poisoning, the triggered performance decreases to 74.8/64.3/80.5, showing that a trigger association learned from only one source class can still affect samples from other semantic categories.
However, the degradation is substantially stronger under multi-class poisoning, where NMI/ARI/ACC fall to 55.7/51.7/73.3.
Compared with the corresponding clean-input performance, multi-class poisoning causes drops of 30.2, 30.4, and 15.1 points in NMI, ARI, and ACC, respectively, whereas the single-class setting produces smaller decreases of 10.9, 17.8, and 8.8 points.

These results suggest that the attack scope plays an important role in determining the severity of the backdoor effect.
When only one semantic class is poisoned, the learned trigger association affects a relatively limited portion of the representation space, and the remaining clean categories continue to provide substantial structural information for clustering.
In contrast, poisoning multiple classes introduces several trigger-associated biases across different semantic regions.
Once these associations are activated at inference time, they interfere with a broader portion of the learned cluster geometry, leading to more severe cross-class confusion and a larger degradation in overall clustering quality.
Therefore, although single-class poisoning already demonstrates that a localized backdoor can generalize beyond its original source class, multi-class poisoning represents a more disruptive threat to the global clustering structure.

\begin{table}[t]
    \centering
    \caption{Comparison between multi-class and single-class poisoning on ImageNet-10.}
    \label{tab:single_vs_multi}
    \setlength{\tabcolsep}{3.5pt}
    \renewcommand{\arraystretch}{1.15}
    \begin{tabular}{C{0.32\columnwidth}|*{3}{C{0.08\columnwidth}}|*{3}{C{0.08\columnwidth}}}
        \toprule
        \midrule
        \multirow{2}{*}{\shortstack{Poisoning\\setting}}
        & \multicolumn{3}{c|}{\rule{0pt}{2.4ex}Clean inputs}
        & \multicolumn{3}{c}{\rule{0pt}{2.4ex}Triggered inputs} \\
        \cline{2-7}
        & \rule{0pt}{2.4ex}NMI & ARI & ACC
        & NMI & ARI & ACC \\
        \midrule
        Multi-class poisoned  & 85.9 & 82.1 & 88.4 & 55.7 & 51.7 & 73.3 \\
        Single-class poisoned & 85.7 & 82.1 & 89.3 & 74.8 & 64.3 & 80.5 \\
        \bottomrule
    \end{tabular}
\end{table}

\subsection{Transferability across Contrastive Clustering Architectures}
\label{sec:cross_model}
To verify that the proposed attack is not specific to CC, we further apply the same poisoning strategy to three representative contrastive clustering architectures: \textbf{VTCC}, \textbf{SACC}, and \textbf{PCLC} \cite{ling2022vtcc,deng2023sacc,zhou2025pclc}.
All experiments are conducted on ImageNet-10 with a poisoning ratio of 10\%.

The results show that HBDCC generalizes across different model families.
VTCC suffers the largest degradation: its NMI/ARI/ACC drop from 85.0/81.7/89.4 on clean evaluation to 23.0/5.2/28.8 after triggered poisoning, indicating that the learned cluster structure is almost unusable under attack.
SACC is also heavily affected, with performance dropping from 87.7/84.3/90.5 to 45.9/40.5/65.6.

Among the three models, PCLC is the most robust, but it is by no means immune: its NMI/ARI/ACC still decrease from 88.5/84.9/90.7 to 71.9/50.6/70.9.

These results suggest that HBDCC targets a vulnerability of \emph{contrastive clustering itself}, rather than exploiting a single model implementation.
At the same time, the relative robustness of PCLC implies that stronger representation learning may offer partial resistance, although it cannot fully eliminate the backdoor effect.

\begin{table}[t]
    \centering
    \caption{Attack transferability across different contrastive clustering models on ImageNet-10 (10\% poisoning).}
    \label{tab:cross_model}
    \setlength{\tabcolsep}{3.5pt}
    \renewcommand{\arraystretch}{1.15}
    \begin{tabular}{C{0.20\columnwidth}|*{3}{C{0.10\columnwidth}}|*{3}{C{0.10\columnwidth}}}
        \toprule
        \midrule
        \multirow{2}{*}{Model} & \multicolumn{3}{c|}{\rule{0pt}{2.4ex}Clean model} & \multicolumn{3}{c}{\rule{0pt}{2.4ex}Poisoned model} \\
        \cline{2-7}
        & \rule{0pt}{2.4ex}NMI & ARI & ACC & NMI & ARI & ACC \\
        \midrule
        VTCC \cite{ling2022vtcc} & 85.0 & 81.7 & 89.4 & 23.0 &  5.2 & 28.8 \\
        SACC \cite{deng2023sacc} & 87.7 & 84.3 & 90.5 & 45.9 & 40.5 & 65.6 \\
        PCLC \cite{zhou2025pclc} & 88.5 & 84.9 & 90.7 & 71.9 & 50.6 & 70.9 \\
        \bottomrule
    \end{tabular}
\end{table}

\subsection{Effect of Trigger Type}
\label{sec:trigger_type}
We next investigate the influence of different trigger patterns.
We perform this experiment on ImageNet-10 under the single-class poisoning setting with a poisoning ratio of 10\%.
Three different HTBA-style triggers are used \cite{saha2020hidden}.

As shown in Table~\ref{tab:trigger_type}, different trigger designs lead to noticeably different attack strengths.
The clean model achieves 85.9/82.2/89.3 in NMI/ARI/ACC.
Under Trigger-1, the performance drops to 71.9/58.4/77.9, indicating a successful but moderate attack.
Trigger-2 is the most destructive, reducing the metrics to 68.9/47.7/73.2.
This pattern is likely to survive augmentation more consistently and is therefore easier for the model to absorb as a stable representation cue.
Trigger-3 is weaker, with the metrics remaining at 76.5/69.7/83.1.

This experiment demonstrates that trigger design is a critical factor in contrastive clustering backdoors.
Even when the poisoning protocol and budget are fixed, changing the trigger pattern shifts ACC by nearly ten points and ARI by more than twenty points.
Defenses therefore cannot rely on a single trigger prior; they must consider a broader trigger design space.

\begin{table}[t]
    \centering
    \caption{Effect of different trigger types on ImageNet-10 (single-class poisoning, 10\% poisoning).}
    \label{tab:trigger_type}
    \renewcommand{\arraystretch}{1.12}
    \begin{tabular}{C{0.38\columnwidth}|*{3}{C{0.14\columnwidth}}}
        \toprule
        \midrule
        Setting & NMI & ARI & ACC \\
        \midrule
        Clean model & 85.9 & 82.2 & 89.3 \\
        Trigger-1   & 71.9 & 58.4 & 77.9 \\
        Trigger-2   & 68.9 & 47.7 & 73.2 \\
        Trigger-3   & 76.5 & 69.7 & 83.1 \\
        \bottomrule
    \end{tabular}
\end{table}

\subsection{Effect of Trigger Size}
\label{sec:trigger_size}

Trigger size is another important factor affecting both attack success and the visual footprint of poisoned samples \cite{gu2019badnets,saha2020hidden,nguyen2021wanet}.
To study this factor, we construct triggers of different sizes (e.g., 40$\times$40, 50$\times$50, and 60$\times$60), poison all classes with triggers of the same size, and test the corresponding poisoned model on clean and triggered test sets under a 10\% poisoning ratio.

\begin{figure}[t]
    \centering
    \includegraphics[width=1\columnwidth]{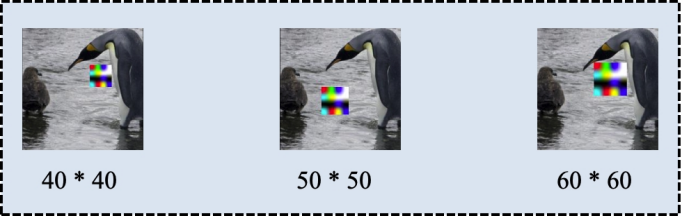}
    \caption{Examples of poisoned samples with different trigger sizes. The trigger size affects both attack effectiveness and visual footprint: smaller triggers are harder for the model to capture, while overly large triggers may interfere with the original image semantics.}
    \label{fig:trigger_size_examples}
\end{figure}

The quantitative results are reported in Table~\ref{tab:trigger_size}.
The clean metrics remain close to the clean model across all tested trigger sizes, indicating that changing the trigger size does not by itself destroy benign clustering utility.
The triggered metrics, however, vary substantially.
The 50$\times$50 trigger produces the strongest attack, reducing NMI/ARI/ACC to 55.7/51.7/73.3, compared with 85.6/81.8/89.0 for the clean model evaluated with the same triggered test protocol.
The 40$\times$40 trigger is less destructive, with triggered performance remaining at 72.2/68.8/82.9, suggesting that a smaller patch is more likely to be weakened by random cropping, resizing, or color perturbation.
The 60$\times$60 trigger is also weaker than the 50$\times$50 trigger, with NMI/ARI/ACC of 70.9/67.0/79.4.
This result shows that increasing the trigger size does not monotonically strengthen the attack: an oversized patch may interfere with the original image semantics instead of acting as a compact and stable shortcut.

\begin{table}[t]
    \centering
    \caption{Effect of trigger size on ImageNet-10 under 10\% poisoning.}
    \label{tab:trigger_size}
    \small
    \setlength{\tabcolsep}{3.8pt}
    \renewcommand{\arraystretch}{1.15}
    \begin{tabular}{C{0.25\columnwidth}|*{3}{C{0.083\columnwidth}}|*{3}{C{0.083\columnwidth}}}
        \toprule
        \midrule
        \multirow{2}{*}{Setting} & \multicolumn{3}{c|}{\rule{0pt}{2.4ex}Clean inputs} & \multicolumn{3}{c}{\rule{0pt}{2.4ex}Triggered inputs} \\
        \cline{2-7}
        & \rule{0pt}{2.4ex}NMI & ARI & ACC & NMI & ARI & ACC \\
        \midrule
        Clean model       & 86.2 & 82.5 & 89.5 & 85.6 & 81.8 & 89.0 \\
        40$\times$40      & 84.9 & 81.2 & 88.6 & 72.2 & 68.8 & 82.9 \\
        50$\times$50      & 85.9 & 82.1 & 88.4 & \textbf{55.7} & \textbf{51.7} & \textbf{73.3} \\
        60$\times$60      & 86.1 & 83.9 & 91.5 & 70.9 & 67.0 & 79.4 \\
        \bottomrule
    \end{tabular}
\end{table}

Overall, an effective trigger needs to occupy a middle range: large enough to survive augmentation and influence the representation, but small enough to avoid overwhelming the clean semantic content.
This trade-off explains why trigger size should be treated as an attack parameter rather than a cosmetic detail.

\begin{table*}[t]
\centering
\caption{Comparison of Trigger Carriers within HBDCC on ImageNet-10
under 10\% Poisoning.}
\label{tab:four_trigger_results}
\small
\setlength{\tabcolsep}{4.5pt}
\renewcommand{\arraystretch}{1.18}
\begin{tabular*}{\textwidth}{@{\extracolsep{\fill}}cl*{6}{c}@{}}
\toprule
\midrule
\multirow{2}{*}{Poisoning Ratio} &
\multirow{2}{*}{Trigger Carrier} &
\multicolumn{3}{c}{Clean inputs} &
\multicolumn{3}{c}{Triggered inputs} \\
\cmidrule(lr){3-5}\cmidrule(l){6-8}
& & NMI & ARI & ACC & NMI & ARI & ACC \\
\midrule
\multirow{5}{*}{10\%}
& Default HBDCC
& 85.9 & 82.1 & 88.4
& \textbf{55.7} & \textbf{51.7} & \textbf{73.3} \\
& Blended~\cite{chen2017targeted}
& 85.1 & 81.2 & 88.3
& 78.3 & 73.5 & 85.6 \\
& SIG-style~\cite{barni2019sig}
& 84.6 & 80.1 & 86.9
& 73.5 & 68.8 & 81.7 \\
& WaNet-style~\cite{nguyen2021wanet}
& 86.0 & 82.8 & 89.5
& 77.1 & 72.2 & 84.0 \\
& FTrojan-style~\cite{wang2022ftrojan}
& 85.8 & 82.1 & 88.9
& 80.2 & 75.7 & 85.8 \\
\bottomrule
\end{tabular*}
\end{table*}

\subsection{Generality across Different Trigger Carriers}
\label{sec:trigger_generality}

\begin{figure}[t]
    \centering
    \includegraphics[width=\columnwidth]{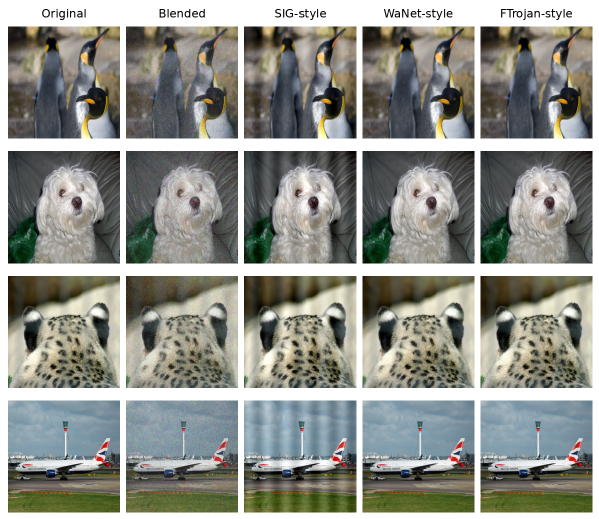}
    \caption{Examples of the four alternative trigger carriers adapted to the HBDCC poisoning protocol: Blended, SIG-style, WaNet-style, and FTrojan-style.}
    \label{fig:alternative_trigger_examples}
\end{figure}

The preceding experiments use the default HBDCC trigger construction.
We now examine whether the observed effect extends to other trigger carriers by adapting Blended~\cite{chen2017targeted}, SIG-style~\cite{barni2019sig}, WaNet-style~\cite{nguyen2021wanet}, and FTrojan-style~\cite{wang2022ftrojan} triggers to the HBDCC poisoning protocol.
These carriers introduce image blending, sinusoidal perturbations, spatial warping, and frequency-domain perturbations, respectively.
We use their trigger constructions within the same unlabeled contrastive clustering pipeline.
All experiments use CC on ImageNet-10 with a poisoning ratio of 10\%.
For each carrier, we independently construct the poisoned training set and the corresponding triggered test set, keeping all other training and evaluation settings unchanged.
Representative examples are shown in Fig.~\ref{fig:alternative_trigger_examples}.

Table~\ref{tab:four_trigger_results} reports the clustering performance on clean and triggered inputs.
The four alternative carriers obtain clean NMI, ARI, and ACC values of 84.6--86.0, 80.1--82.8, and 86.9--89.5, respectively.
These values are close to the corresponding results of 85.9/82.1/88.4 obtained with the default trigger, indicating similar clean-data utility across the evaluated carriers.

All four alternative carriers yield lower clustering scores on triggered inputs than on their corresponding clean inputs.
Among these alternatives, SIG-style produces the lowest triggered NMI/ARI/ACC of 73.5/68.8/81.7, followed by WaNet-style at 77.1/72.2/84.0, Blended at 78.3/73.5/85.6, and FTrojan-style at 80.2/75.7/85.8.
Relative to each model's own clean-input performance, the alternatives yield decreases of 5.6--11.1 points in NMI, 6.4--11.3 points in ARI, and 2.7--5.5 points in ACC.
Thus, the degradation is consistent across carriers, although its magnitude varies by carrier and metric.

The default HBDCC trigger produces the strongest disruption under the evaluated setting, with triggered NMI/ARI/ACC of 55.7/51.7/73.3.
These results correspond to decreases of 30.2, 30.4, and 15.1 points from its clean-input performance, substantially larger than those of the four alternatives.
The comparison therefore supports the flexibility of the HBDCC poisoning protocol across multiple trigger forms, while showing that the default construction remains more effective in this experiment.

The carrier-dependent results are consistent with our dual-level analysis: trigger construction may affect how reliably a non-semantic cue is retained across augmented views and subsequently reinforced through instance-level representation learning and cluster-level consistency.
This interpretation suggests a possible explanation for the observed differences; the carrier comparison alone does not isolate the contribution of either learning level.

\subsection{Knowledge-Distillation Defense}
\label{sec:defense}
Finally, we evaluate a defense strategy based on \textbf{knowledge distillation} \cite{hinton2015distilling,saha2022sslbackdoor}.
The core idea is to use a small amount of clean unlabeled data and a clean teacher signal to guide a student model away from trigger-dependent representations, thereby weakening the association between trigger patterns and clustering semantics.
In our implementation, the clean teacher is trained without poisoned samples, the student is initialized from the attacked model, and distillation uses a manually screened clean unlabeled subset covering about 10\% of the dataset.
The distillation process runs for 50 epochs.

The results demonstrate that this strategy is effective but not complete.
On the triggered ImageNet-10 test set, the attacked model obtains 55.7/51.7/73.3 in NMI/ARI/ACC before defense.
After knowledge distillation, the metrics recover to 78.6/74.8/85.8, corresponding to gains of 22.9 NMI, 23.1 ARI, and 12.5 ACC points.
The larger recovery in NMI and ARI suggests that distillation helps restore global cluster consistency, while the remaining ACC gap indicates that some triggered samples are still assigned to incorrect clusters.
This is consistent with the mechanism of the attack: once the trigger is embedded into the representation space, removing it requires more than a small amount of clean-data rehearsal.
Therefore, knowledge distillation should be viewed as an effective \emph{mitigation} strategy rather than a complete solution.

\begin{table}[t]
    \centering
    \caption{Performance before and after knowledge-distillation defense on the triggered ImageNet-10 test set.}
    \label{tab:defense}
    \renewcommand{\arraystretch}{1.12}
    \begin{tabular}{C{0.42\columnwidth}|*{3}{C{0.13\columnwidth}}}
        \toprule
        \midrule
        Defense setting & NMI & ARI & ACC \\
        \midrule
        Before defense      & 55.7 & 51.7 & 73.3 \\
        After distillation  & 78.6 & 74.8 & 85.8 \\
        \bottomrule
    \end{tabular}
\end{table}

\subsection{Discussion}
\label{sec:discussion}
The experiments support four conclusions.
First, HBDCC satisfies the defining property of a backdoor attack: the poisoned model usually keeps plausible clean-data behavior, but its clustering quality drops markedly once the trigger is present.
Second, the attack can be effective with a modest poisoning budget, but its strength depends on dataset separability; fine-grained or visually diverse datasets are easier to disrupt.
Third, the vulnerability is not restricted to one architecture, because VTCC, SACC, and PCLC all suffer clear degradation under the same poisoning protocol.
Fourth, trigger design and attack scope matter in different ways: the trigger pattern controls how strongly the cue survives augmentation, while the attack scope determines how broadly the learned association affects cluster geometry.
Although knowledge distillation alleviates the attack substantially, the residual performance gap indicates that defending deep contrastive clustering against backdoors remains an open problem \cite{tejankar2023patchsearch,zhang2024corruptencoder,chen2025noisyalignment}.

\section{Conclusion}
\label{sec:conclusion}

This paper studies a training-time backdoor threat against deep contrastive clustering.
Our results show that augmentation-persistent triggers can be learned as invariant cues by the instance-level objective and reinforced by cluster-level consistency, implanting a backdoor in the representation geometry without label supervision or changes to the victim pipeline.
We propose \textbf{HBDCC}, a data-only attack that poisons a small subset of unlabeled training images.
Across representative contrastive clustering models, triggered inputs substantially degrade clustering performance while clean-data utility is generally preserved; the effect varies by dataset and is particularly strong in challenging fine-grained settings.
Our analyses further show that attack strength depends on poisoning scope and trigger design, and that knowledge distillation with clean unlabeled data can mitigate, but not fully remove, the backdoor.
These findings highlight the need to treat the security of unsupervised and self-supervised clustering systems as an important complement to clustering accuracy.

\textbf{Limitations.}
This study is limited to representative image benchmarks and contrastive clustering architectures; larger datasets, non-visual domains, and foundation-model-based clustering remain to be investigated.
Our attack uses manually designed trigger banks, so the transferability and concealment of adaptive or learned triggers are open questions.
Future work should develop clustering-specific detection and robust training methods, including trigger-invariant representation learning, robust assignment regularization, and poisoned-sample screening for unlabeled data \cite{tejankar2023patchsearch,zhang2024corruptencoder,zhang2025invisible,chen2025noisyalignment,wang2019neuralcleanse,zheng2022preadistribution}.
Since this work is dual-use, any release of attack details or code should include defensive guidance and evaluation safeguards.

\ifCLASSOPTIONcompsoc
\section*{Acknowledgments}
\else
\section*{Acknowledgment}
\fi
This work was supported in part by National Natural Science Foundation of China (No.62272198), 
in part by the Special Funds for the Cultivation of Guangdong College Students’ Scientific and Technological Innovation (Climbing Program Special Funds) under Grant pdjh2025ak028, 
and in part by the Excellent Graduate Student Cultivation Program of Jinan University (Grant No.2026CXB080).
\ifCLASSOPTIONcaptionsoff
\newpage
\fi
\bibliographystyle{IEEEtran}

\bibliography{bdcc}

\end{document}